\documentclass{aa}
\usepackage{times,graphics,rotate,psfig,epsfig}
\begin{document}
\title{Fundamental parameters of Galactic luminous OB stars\\
VI. Temperatures, masses and WLR of Cyg OB2 supergiants\thanks{The INT 
is operated
in the island of La Palma by the ING in the Spanish Observatorio
de El Roque de los Muchachos of the Instituto de Astrof\'{\i}sica de
Canarias}}
\author{A. ~Herrero\inst{1,2}, J. Puls\inst{3}, F. Najarro\inst{4}}
\offprints{A. ~Herrero, \email{ahd@ll.iac.es}}
\mail{ahd@ll.iac.es}
\institute{Instituto de Astrof\'\i sica de Canarias, E-38200 La Laguna, 
Tenerife, Spain
\and
Departamento de Astrof\'{\i}sica, Universidad de La Laguna,
Avda. Astrof\'{\i}sico Francisco S\'anchez, s/n, E-38071 La Laguna, Spain
\and
Universit\"ats-Sternwarte M\"unchen, Scheinerstr. 1, D-81679 M\"unchen, 
Germany
\and
Instituto de Estructura de la Materia, Consejo Superior de Investigaciones
Cient\'{\i}ficas, C/ Serrano 121, E-28006 Madrid, Spain}

\date{Received 3 June 2002; accepted 30 September 2002 }

\titlerunning{Temperatures, masses and WLR of Cyg OB2 supergiants}
\authorrunning{Herrero, Puls, Najarro}
\abstract{
We have analyzed six OB supergiants and one giant covering spectral types
from O3 to B1 in the Galactic OB
association Cyg OB2 by means of an updated version of FASTWIND
(\cite{sph97}) that includes an approximate treatment of metal line
blocking and blanketing. This large coverage in spectral type allows us
to derive a new temperature scale for Galactic O supergiants that is
lower than the one obtained by using pure H--He models,
either plane-parallel and hydrostatic or spherical with mass-loss. The lower
temperatures are thus a combined effect of line blanketing and the large
mass-loss rates. In some cases, the newly derived effective temperature is
reduced by up to 8\,000 K. Changes are larger for earlier stars with large
mass--loss rates. As a consequence, luminosities are modified as well,
which results in a lower number of emerging ionizing photons and
reduces the mass discrepancy. Although there are still significant
differences between spectroscopic and evolutionary masses, we do not find
any obvious systematic pattern of those differences. We derive mass--loss
rates and the corresponding wind momentum--luminosity relation for the
analyzed stars. Although consistent with previous results by
\cite{puls96} for Galactic stars, our relation is better defined due to a
reduction of errors related to stellar distances and points to a possible
separation between extreme Of stars (Of$^+$, Of$^*$) and stars with more
moderate morphologies. However this finding is only tentative, as the
statistics are still scarce.

}
\maketitle

\section{Introduction}
New winds are blowing from hot massive stars. The development of new model
atmosphere codes during recent years (\cite{hub95}; \cite{sph97};
\cite{hill98}; \cite{paul01}) employing improved numerical techniques and
detailed atomic models offers the opportunity to derive more realistic
stellar parameters that can be used for the research of astrophysical
problems. Thus very recently Martins et al. (2002)
have presented a new
temperature scale for massive O dwarfs that lowers considerably the previous
one given by Vacca et al. (1996)
as a result of strong metal line blanketing. The
effect should be even larger for supergiants, where strong mass--loss should
add to metal line opacity to yield even smaller effective temperatures.
First indications of this behaviour were obtained for example by
Crowther \& Bohannan (1997), Herrero et al. (2000) or Fullerton et al. (2000). 
More recently, Crowther et al. (2002)
have presented an analysis of 4 supergiants in the LMC and the SMC with
similar trends and Bianchi \& Garc\'{\i}a (2002)
obtain comparable results for O6--O7 stars in the Milky Way.

The stellar masses of O stars are still a subject of debate. A large part of
the uncertainty originates from the so-called {\it mass-discrepancy}
(Herrero et al., 1992): masses derived from a spectroscopic analysis using
hydrostatic, plane--parallel model atmospheres are systematically lower than
those predicted by non--rotating evolutionary models. Unified model
atmospheres are expected to reduce the discrepancy, especially 
if new evolutionary models account for rotationally induced
mixing in stars (\cite{h00};
\cite{mey00}). To this end, considerable effort has been made in the last
decade to account for the possibility of He enriched and CNO contaminated
atmospheres in early type stars. Efforts to improve the derived
abundances have lowered them, but without completely ruling out He
enhancements (\cite{villa02}).

On the other hand, the presence of a wind momentum-luminosity relation (WLR) for
hot, massive stars offers the unique opportunity to derive stellar distances
directly from the analysis of the observed spectra with an accuracy of about
10$\%$ (Kudritzki et al., 1999). For this purpose, the coefficients in the
relation
$$
log (\dot M V_\infty R^{0.5}) \approx \frac{1}{\alpha^\prime} log L + C =
x~ log (L/L_\odot) + log D_0
$$
have to be calibrated as function of metallicity and spectral type. For
O-stars, this has been done by Puls et al. (1996) for Galactic and MC objects.
However, their WLR shows a large scatter, especially for non-supergiants. It
is reasonable to assume that part of this scatter results from uncertainties
in the individual stellar distances, as their sample was mainly taken from
Herrero et al. (1992), aiming at targets with large apparent brightness.

Following this idea, we started a programme to calibrate the WLR using a
more homogeneous sample of OB supergiants, all of them belonging to the same
OB association. From this approach we expected to reduce the scatter in the
derived WLR. The OB association chosen was Cyg OB2, and the optical
observations and first analysis using plane--parallel, hydrostatic models
were presented in Herrero et al. (1999). 
The required wind terminal velocities were
obtained from HST STIS observations and are discussed in Herrero et al. (2001).

By means of spherical, mass--losing models we will derive here an improved
WLR for those Cyg OB2 supergiants that were observed with the HST. (One of
the stars, Cyg OB2 \#4, is actually a luminosity class III object, but we will
refer to our sample as Cyg OB2 supergiants. When required, we will emphasize
the difference).

This effort is being complemented in our group by other studies to calibrate
the WLR in the Local Group (see McCarthy et al., 1997, Urbaneja et al.,
2002, Bresolin et al., 2002 for M31 and M33).

The remainder of this paper is organized as follows. In Sect.~\ref{code} we
describe the characteristics of the code we have used, and in Sect.~\ref{test} we
present an analysis of the O9 V star 10 Lac as a test case. The analyses of
our targets is given in Sect.~\ref{anal},  and we end with a discussion of
our results in Sect.~\ref{disc} and the conclusions in  Sect.~\ref{conc}.

\section{The code}
\label{code}

The analyses presented here have been performed by means of the latest version
of FASTWIND (an acronym for Fast Analysis of STellar atmospheres with
WINDs), a code which was originally described by Santolaya--Rey et al. (1997)
and was recently updated to include an approximate treatment of 
metal line opacity effects, i.e., metal line blocking and blanketing.

The code follows the philosophy of using suitable physical approximations
allowing for a fast computational time, thus enabling us to calculate a
large number of models (which are necessary to analyze stellar atmospheres
with winds), while being realistic enough to preserve their value as a tool
for determining stellar parameters.

It assumes a $\beta$--velocity law in the wind and calculates a consistent
photospheric structure; the temperature structure is approximated using
``NLTE Hopf functions'' as described in Santoloaya--Rey et al. (1997);
the coupling between
the radiation field and the rate equations is treated within an ALI scheme,
using local ALOs following Puls (1991).
The program allows for a solution
of the rate equations using either Sobolev or comoving frame calculations,
but all calculations presented here were done using comoving frame. The
formal solution to calculate the emergent profiles utilizes a radial
micro-grid to account for the different scales involved and includes Stark
broadening, which is a prerequesite for the analysis of O stars by means of
H and He lines.

The approximate treatment of metal line blocking/blanketing will be
described in detail by Puls (2002),
in the following we will give only a
brief summary. The basic philosophy to calculate the required NLTE metal
opacities (bound-bound, bound-free and free-free) follows the line of
reasoning given by Abbott \& Lucy (1985), Schmutz (1991), 
Schaerer \& Schmutz (1994) and Puls et
al. (\cite{Pulsetal00}), however applying significant improvements. In
particular and most important for realistic results, we have reformulated
the equations of approximate ionization equilibrium (e.g., Puls et al.
\cite{Pulsetal00}, Eq. 43) to account for the actual radiation field (as
function of depth) at all ionization edges (including those from excited
levels) and employ a consistent coupling of rate-equations and mean
intensity, in a way similar to the ALI approach, to avoid any kind of Lambda
Iteration.

The underlying atomic data base has been described by Pauldrach et al.
(1998). In order to save computational effort, the
resulting metal line opacities are averaged in a suitable way (mean of
inverse opacties, in analogy to Rosseland means) over a frequency interval
of order wind terminal velocity before the radiation transport is performed.
Finally, flux conservation (and thus line blanketing) is obtained by
adapting the NLTE-Hopf parameters in a consistent way. The method has been
carefully tested by comparing the results with up-to-date methods, in
particular with calculations performed with TLUSTY (\cite{hub95}) for the
case of dwarfs (see also the next section) and with WMBasic (\cite{paul01})
for the case of stars with winds.

\section{A test case: 10 Lac}
\label{test}

Before applying the modified code to the analysis of our Cyg OB2 sample, we
have analyzed the O9V star 10 Lac as a test case. This star is well suited
for calibration purposes because it is a luminosity class V star with a low
projected rotational velocity, has already been used as standard by
Herrero et al. (1992) and was later considered by 
Hubeny et al. (1998) to study the effects
of line blanketing in plane-parallel, hydrostatic atmospheres.

\begin{figure*}
{\includegraphics{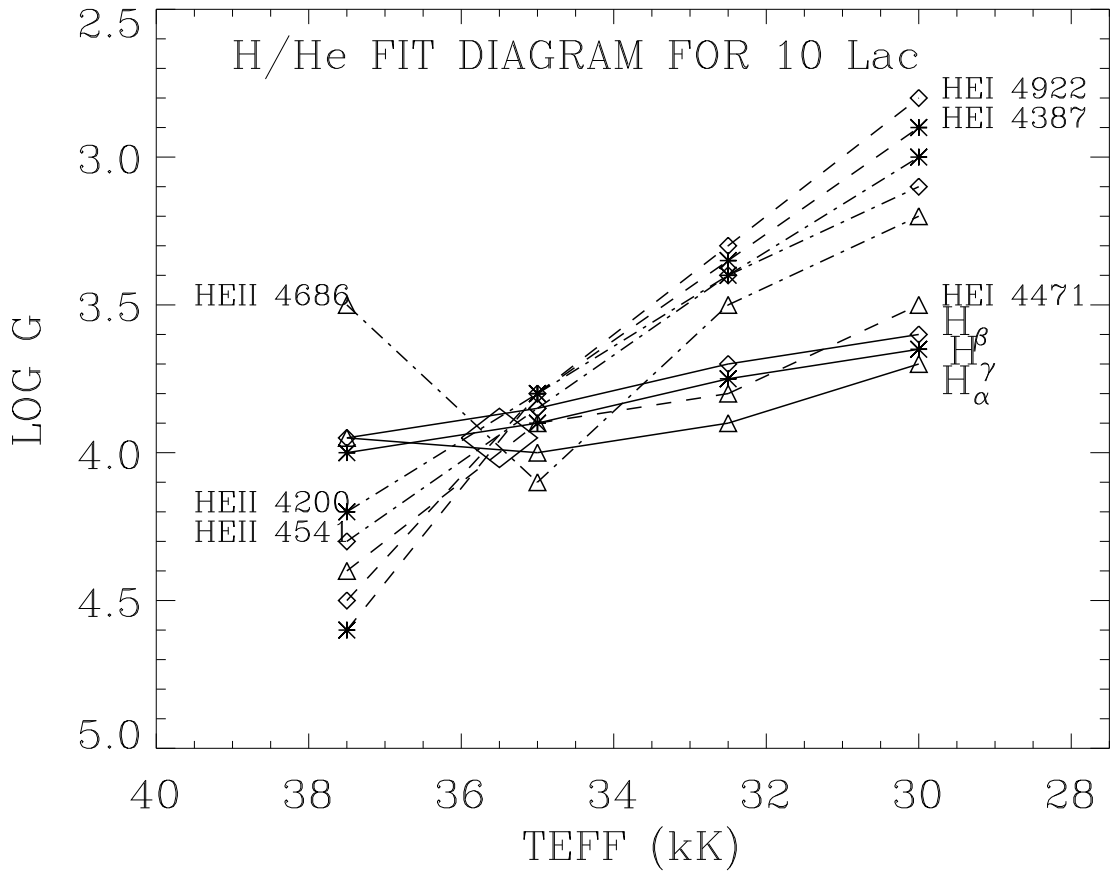}}
\caption[]{Fit diagram to the H/He spectrum of 10 Lac using FASTWIND 
plus approximated line blocking/blanketing. The filled circle marks the
location of the chosen model, and the box around gives the
size of the adopted error box}
\label{10lacdiag}
\end{figure*}

Fig.~\ref{10lacdiag} gives the fit diagram for the H/He spectrum of 10 Lac.
It is very similar to the corresponding fit diagram in Herrero et al.
(1992), but now
it is centered at a lower effective temperature (and a slightly lower
gravity). A comparison of both diagrams also reveals that the dispersion
around the final model is now smaller. Therefore, the present error box has a
width of only 1\,000 K instead of 2\,000 K, as the one in Herrero et
al. (1992). It is
reassuring that all lines, including \ion{He}{ii} $\lambda$
4200 that could not be fitted by Herrero et al. (1992),
lie within the error box.

The stellar parameters of 10 Lac we have derived here are
$T_{\rm eff}$= 35\,500$\pm$500 K, $\log g$= 3.95$\pm$0.10 and 
$\epsilon$=$\frac{\rm{N(He)}}{\rm {N(H)+N(He)} }$= 0.09$\pm$0.03, N(X) 
being the abundance of element X by number.
The only significant difference compared to the results from
Herrero et al. (1992) is the effective temperature, now lower by 2\,000 K.
This is in complete agreement with the result obtained by
Hubeny et al. (1998). These authors estimated a temperature of
35\,000 K for 10 Lac using TLUSTY, a plane-parallel, hydrostatic,
line--blanketed model. 

\begin{figure*}
{\includegraphics{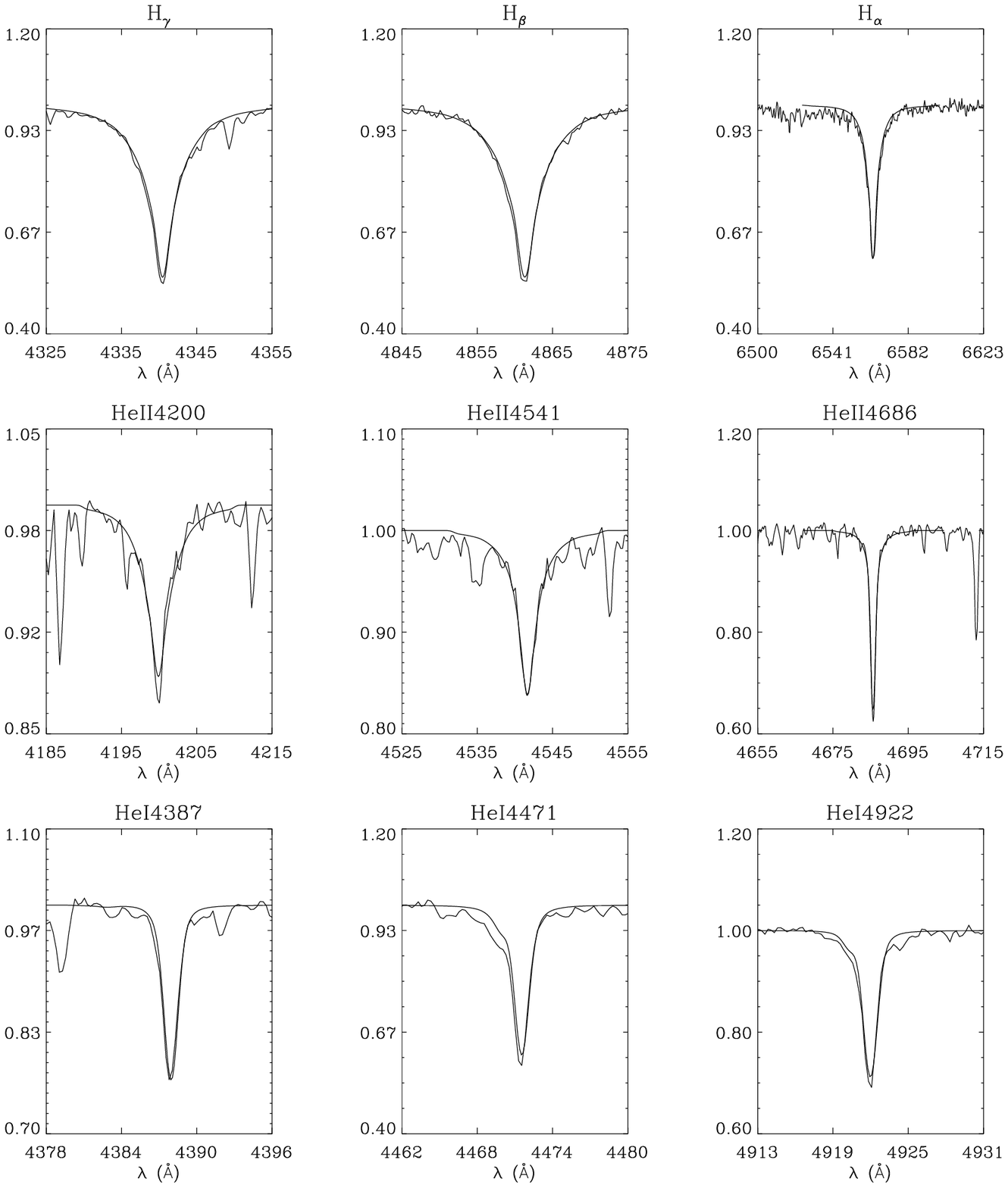}}
\caption[]{The fit to the H/He spectrum of 10 Lac using FASTWIND with
approximated line blocking/blanketing. The ordinate gives the
relative flux values. Note that the scales are different for
each line. See text for comments.}
\label{10lac}
\end{figure*}

Fit diagrams have the drawback that they only give the best possible fit
for the chosen constraints. They rely on interpolations and sometimes 
(when using EWs) do not account for the profile shape.
The actual final fit may still be poor.
Fig.~\ref{10lac} shows the line spectra for
our final model of 10 Lac. Good agreement is found for all lines, although a
few details are not perfectly reproduced. In particular, the core of
\ion{He}{ii} $\lambda$ 4200 is too weak, which is  also true for the forbidden
component of \ion{He}{i} $\lambda$ 4471. Besides this, however, the final fit
is perfectly consistent in all other aspects.

Our result also agrees with the study of Martins et al. (2002), who recently
found that using pure H/He models (as \cite{h92} did) results in an
effective temperature scale for O dwarfs hotter by 1\,500 to 4\,000 K 
compared to using line--blanketed models.

These authors have derived a new effective temperature scale for O dwarfs
using CMFGEN (\cite{hill98}), a spherical code including mass--loss and
blanketing. In their $T_{\rm eff}$\, scale, O9V stars are located at 33\,000 K.
However, their scale is calibrated using the 
equivalent width (EW) -- spectral
classification relations of Conti \& Alschuler (1971) and Mathys (1998), 
and thus we have to compare our result with Martins et al. 
using the classification of 10 Lac in this system.

Conti \& Alschuler (1971) have classified 10 Lac as O8 III and not as a
luminosity class V star (which have been considered by \cite{martins02}),
although the star lies just at the border between both luminosity
classes. The luminosity class III is mainly due to the low EW quoted
by Conti \& Alschuler (1971) for the \ion{He}{i} $\lambda$4143 line.
Other EWs quoted by these authors are consistent with our observations
which show a much larger value for this line, resulting in a luminosity
class V within their classification scheme. Thus, we conclude
that 10 Lac should be classified as O8 V in the system
of Conti \& Alschuler (1971)
(whatever the reason for the low EW in \cite{conti71} was).

The effective temperature in the Martins et al. scale for O8 dwarfs
lies between 36\,000 and 35\,000 K, in perfect agreement with our result.
Therefore we regard our result as fully consistent with recent
findings using more elaborate but also more expensive fully blanketed NLTE
model atmospheres.

Nevertheless we have analyzed 10 Lac with our version of CMFGEN that
includes a consistent photospheric structure (\cite{naj02}) and have fitted
the line profiles instead of only using their EWs. Our results from CMFGEN
completely agree with those from FASTWIND.

\begin{figure*}
{\includegraphics[width=12cm]{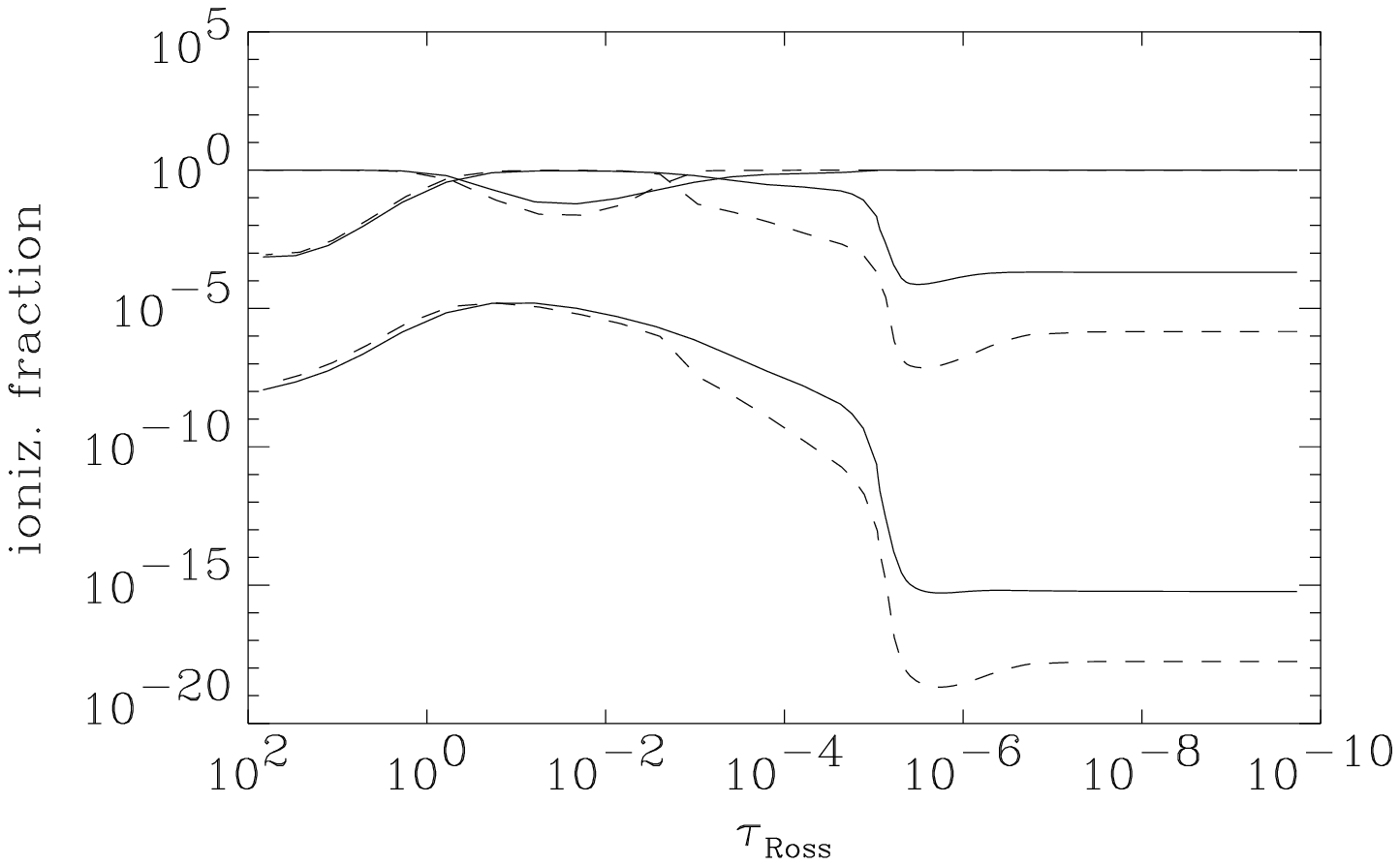}}
\caption[]{The He ionization fractions in two models with $T_{\rm eff}$= 35\,500 K,
$\log g$= 3.95 and $\epsilon$= 0.09, with approximated line--blocking/blanketing 
(solid line)
and without (dashed line). At typical photospheric line formation depths the
ionization degree of the model including metal opacities is larger. Thus,
lower temperatures are required in these models to reproduce the He
spectrum.}
\label{ionization}
\end{figure*}

The reason for the lower temperatures derived is twofold (see also
\cite{martins02}). On the one hand, the radiation field
which is backscattered due to the additional opacity produces a larger
(E)UV radiation field. On the other hand, due to line-blanketing the electron
temperature rises in photospheric regions. Both effects favour a higher
ionization degree at lower effective temperatures, compared to
unblanketed models. This effect can be clearly seen in
Fig.~\ref{ionization}.

The analysis of 10 Lac gives us an indication of what we can expect when
introducing metal line opacity (namely lower effective temperatures), either
in the form of traditional line--blanketing (as \cite{hub98},
\cite{martins02} and references therein) or including the
line--blanketing via adapted Hopf--parameters (as here). Although with our
method we do not {\it force} flux conservations, in all models calculated
here the flux is conserved to better than 3$\%$ at all depths, where the
remaining deviations have no impact on the emergent fluxes and profiles.

Our analysis of 10 Lac also gives us an idea of the error bar we can expect
for the stellar parameters. For a resolution of 8\,000, a S/N of 200 and a
projected rotational velocity of 40 km s$^{-1}$, the estimated errors are
$\pm$500 K in $T_{\rm eff}$, $\pm$0.1 dex in $\log g$\, and $\pm$0.03 in $\epsilon$. For what
follows we shall note here that this analysis does not give us information
about the mass loss rate or the $\beta$ exponent in the wind velocity law,
as the wind of 10 Lac is negligible. We obtain an upper limit of 10$^{-8}$
${\rm M}_{\odot}$ yr$^{-1}$, but the fit has the same quality for any value below that
limit. (The fit presented here was performed with \.M= 10$^{-10}$ ${\rm M}_{\odot}$\,
yr$^{-1}$).

\section{Analysis of the Cyg OB2 supergiants}
\label{anal}

We make use of optical and UV spectra that have been presented elsewhere
(\cite{h99}; \cite{h01}). The optical spectra have been newly rectified,
resulting in a less pronounced bump in the 4630-4700 \AA~ region, which,
however, does not affect our present analysis. Basic data, together with
previously determined parameters adopted here are listed in
Table~\ref{previous}, while the new parameters determined in this work are
provided in Table~\ref{finpar}. Note that the gravities in that table have
been corrected for the effect of centrifugal acceleration (in the same way
as in \cite{h92}). 
This effect is small (modifying only the last digit of the entries in 
our table) and the gravities
actually used in our calculations are always given
by the nearest lower value ending with  ``0'' or ``5'' in
the second decimal (i.e., a corrected value of 3.52 corresponds
to a model value of 3.50). The corrected gravities have been used
for the calculation of stellar masses.

Errors adopted for the parameters given in Table~\ref{finpar} have been taken
from the errors obtained for Cyg OB2 \#7 (see below) or from those for 10
Lac, depending on the stellar parameters, the spectrum quality and the fit
conditions. A summary of all errors is given in Table~\ref{errors}.

For Cyg OB2 \#7, errors have been estimated ``by eye'' from a microgrid of
models around the final one. At fixed $\beta$ (the exponent of the
velocity law), this resulted in $\pm$1000 K in $T_{\rm eff}$, $^{+0.15}_{-0.10}$ dex
in $\log g$, $\pm$0.05 dex in $\epsilon$\, and $^{+0.05}_{-0.10}$ dex in \.M. $\beta$ is
determined from the form of the H$_{\rm \alpha}$\, wings, and its uncertainty is
estimated to be $\pm$0.10. This has an influence on the derived effective
temperature and mass--loss rates. Therefore, the adopted error for $T_{\rm eff}$~ has
been increased to $\pm$1500 K and that for \.M to $^{+0.10}_{-0.15}$.

The errors for radii and masses depend on the error in the absolute
magnitude. This is assumed to be $\pm$0.1 from the work by 
Massey \& Thompson (1991). 
(The influence of the error in $T_{\rm eff}$ on the stellar radius
is marginal and has been neglected here). For the analysis we have adopted a
microturbulence of 10 km s$^{-1}$. Tests indicate that this parameter is of
no relevance for the results presented here.

In the following we will comment on the individual analyses. Further
discussions about the individual stars were presented in Herrero et al. 
(2001).

\begin{table*}
  \caption[]{Cyg OB2 supergiants studied in this work. All numerical
identifications are taken from Schulte (1958). Magnitudes have been adopted
from Massey \& Thompson (1991), 
as well as spectral types, except for Cyg OB2 $\#$11
and $\#$4, that are taken from Walborn (1973). 
S/N values have
been estimated from the rms of the continuum at several wavelength intervals.
Velocities are given in km s$^{-1}$.}

\label{obs}
    \begin{tabular}{r r c r r r c}
      \hline 
Ident& $V$& Spectral& $M_{\rm v}$ & S/N & $V_{\rm r}${\thinspace}sin{\thinspace}$i$ & $v_{\rm \infty}$ \\
     & mag.& Type   &     &     &     &     \\
      \hline 

  7 & 10.55 & O3 If       & -5.91 & 140 & 105 & 3080 \\
 11 & 10.03 & O5 If$^+$   & -6.51 & 190 & 120 & 2300 \\
 8C & 10.19 & O5 If       & -5.61 & 195 & 145 & 2650 \\
 8A &  9.06 & O5.5 I(f)   & -7.09 & 135 &  95 & 2650 \\
  4 & 10.23 & O7 III((f)) & -5.44 & 230 & 125 & 2550 \\
 10 &  9.88 & O9.5 I      & -6.86 & 145 &  85 & 1650 \\
  2 & 10.64 & B1 I        & -4.64 & 195 &  50 & 1250 \\
 \end{tabular}
\label{previous}
\end{table*}

\begin{table*}
\caption{Results obtained using FASTWIND plus line blocking/blanketing.
Effective temperatures have been derived from the He ionization equilibrium.
Gravities include the correction for centrifugal acceleration. $\epsilon$~ is the
He abundance by number relative to H plus He; R, L and M are given in solar
units. $M_{\rm s}$, $M_{\rm ev}$~ and $M_{\rm 0}$~ are the spectroscopic, present evolutionary and
initial evolutionary masses, the two latter from the models by Schaller et
al. (1992). $\dot M$ is given in solar masses per year, and $\log$MWM means
the logarithm of the Modified Wind Momentum rate, \.M$v_{\rm \infty}$ R$^{0.5}$}
    \begin{tabular}{r l c c c c c c c r c c c}
      \hline 
Ident& Spectral& $T_{\rm eff}$ & $\log g$ & $\epsilon$ & R & $\beta$ & $\dot M$ & $\log$L & $M_{\rm s}$ & $M_{\rm ev}$ & $M_{\rm 0}$ & $\log$MWM\\
     &  Type   &       &    &      &   &         &          &         &      & & &     \\
      \hline 

  7 & O3 If$^*$   & 45.5 & 3.71 & 0.23 & 14.6 & 0.90 & 9.86e-6 & 5.91 & 39.7 & 67.4 & 69. & 29.864 \\
 11 & O5 If$^+$   & 37.0 & 3.61 & 0.09 & 22.2 & 0.90 & 9.88e-6 & 5.92 & 73.0 & 58.1 & 63. & 29.829 \\
 8C & O5 If       & 41.0 & 3.81 & 0.08 & 13.3 & 0.90 & 2.25e-6 & 5.66 & 42.2 & 46.1 & 48. & 29.137\\
 8A & O5.5 I(f)   & 38.5 & 3.51 & 0.09 & 27.9 & 0.70 & 1.35e-5 & 6.19 & 90.5 & 78.4 & 95. & 30.076 \\
  4 & O7 III((f)) & 35.5 & 3.52 & 0.09 & 13.6 & 1.00 & 8.60e-7 & 5.41 & 21.8 & 32.6 & 34. & 28.708 \\
 10 & O9.5 I      & 29.0 & 3.11 & 0.09 & 30.3 & 1.00 & 3.09e-6 & 5.77 & 43.1 & 43.7 & 48. & 29.248\\
  2 & B1 I        & 28.0 & 3.21 & 0.09 & 11.3 & 1.00 & 6.92e-8 & 4.85 &  7.5 & 18.1 & 19. & 27.263\\
 \end{tabular}
\label{finpar}
\end{table*}

\begin{table*}
\caption{Errors adopted for the parameters given in Table~\ref{finpar},
in the same units as in that table. If only one value is provided, it should
be preceeded by the $\pm$ sign. See text for comments.}
    \begin{tabular}{r l r c c c c c c r c c c}
      \hline 
Ident& Spectral& $\Delta$ & $\Delta$  & $\Delta$  & $\Delta$  & $\Delta$  & $\Delta$ & $\Delta$  & $\Delta$  & $\Delta$  & $\Delta$ & $\Delta$ \\
     &   Type  & $T_{\rm eff}$    & $\log g$ & $\epsilon$ & $\log$R & $\beta$ & $\log \dot M$ & $\log$L & $M_{\rm s}$ & $M_{\rm ev}$ & $M_{\rm 0}$ & $\log$MWM \\
      \hline 

  7 & O3 If$^*$   & 1.5 & $^{+0.15}_{-0.10}$ & $^{+0.10}_{-0.05}$ & 0.02 & 0.10& $^{+0.10}_{-0.15}$ & 0.10 & $^{+17}_{-13}$ & $^{+11}_{-8}$ & $^{+11}_{-8}$ & $^{+0.12}_{-0.17}$ \\
 11 & O5 If$^+$   & 1.5 & $^{+0.15}_{-0.10}$ & 0.03 & 0.02 & 0.10 & $^{+0.10}_{-0.15}$ & 0.11 & $^{+32}_{-24}$ & $^{+10}_{-9}$ & $^{+11}_{-10}$ & $^{+0.12}_{-0.17}$\\
 8C & O5 If       & 1.5 & 0.10 & 0.03 & 0.02 & 0.10 & $^{+0.10}_{-0.15}$ & 0.10 & 14 & 5 & 5 & $^{+0.12}_{-0.17}$\\
 8A & O5.5 I(f)   & 1.5 & 0.10 & 0.03 & 0.02 & 0.10 & $^{+0.10}_{-0.15}$ & 0.11 & 29 & $^{+4}_{-7}$ & 15 & $^{+0.12}_{-0.17}$\\
  4 & O7 III((f)) & 1.0 & 0.10 & 0.03 & 0.02 & 0.10 & 0.10 & 0.09 & 7 & 4 & 5 & 0.12\\
 10 & O9.5 I      & 1.0 & 0.10 & 0.03 & 0.02 & 0.10 & 0.10 & 0.10 & 14 & $^{+12}_{-6}$ & $^{+16}_{-6}$ & 0.13\\
  2 & B1 I        & 1.0 & 0.10 & 0.03 & 0.04 & 0.10 & 0.10 & 0.14 & 3 & 3 & 3 & 0.13\\
 \end{tabular}
\label{errors}
\end{table*}

\paragraph{Cyg OB2 \#7}

\begin{figure*}
{\includegraphics{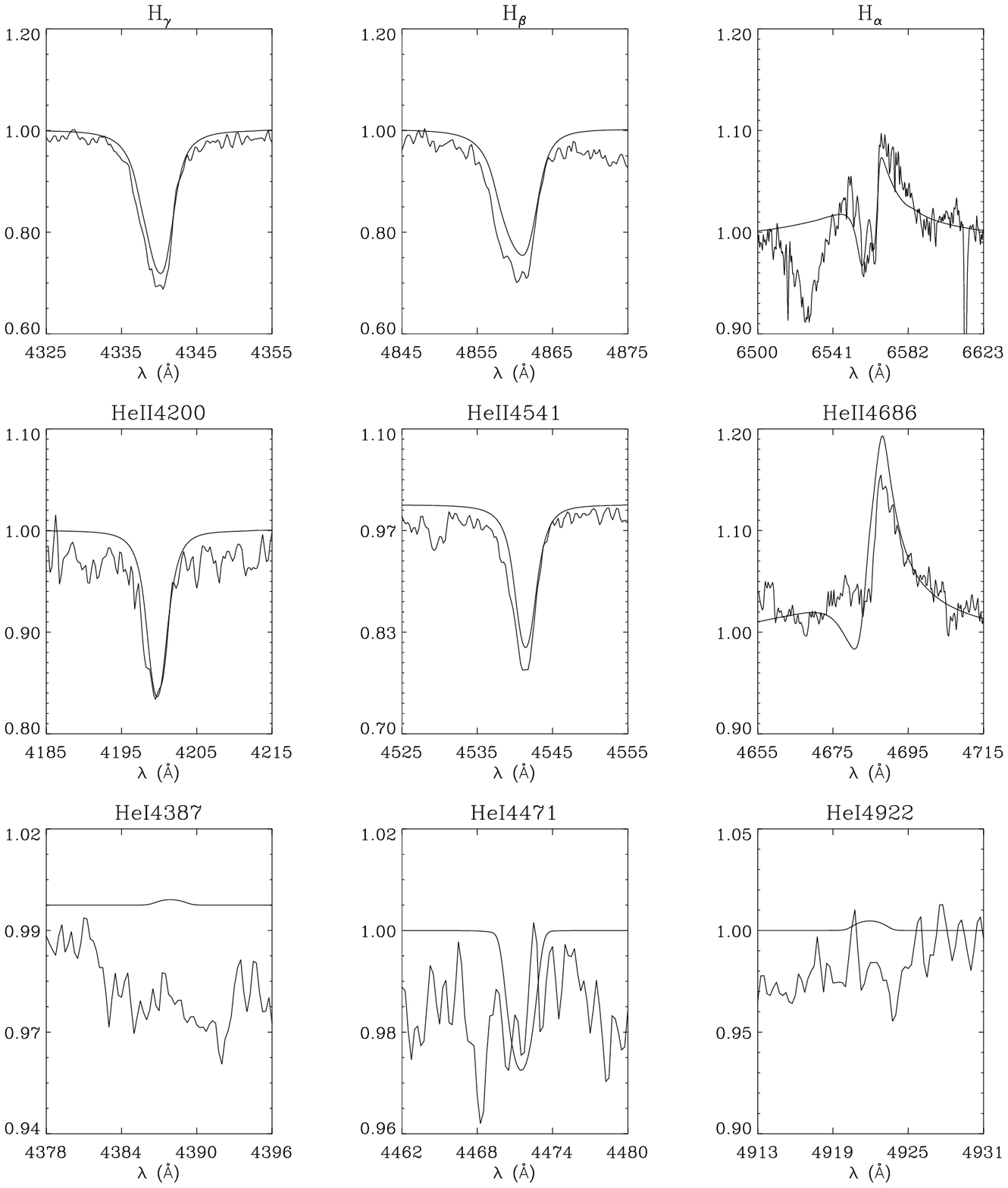}}
\caption[]{As Fig.~\ref{10lac}, however for CygOB2 \#7. See text for comments}
\label{fit7}
\end{figure*}

The final fit to the observed H/He spectrum of Cyg OB2 \#7 is shown in
Fig.~\ref{fit7}. The good fit to the H$_{\rm \alpha}$\, profile is accompanied by a much
poorer fit to the other two Balmer lines (and also to H$_{\rm \delta}$, not displayed
here). This behaviour reproduces the one found in 
Herrero et al. (2000): for stars with
strong winds we could not obtain a consistent fit for all Balmer lines at a
given mass--loss rate. The situation has improved with the new version of
our code, but the discrepancy still reaches a 30$\%$ effect, by which the
mass--loss rate has to be reduced (from 10$^{-5}$ to
7.7$\times$10$^{-6}$) in order to fit H$_{\rm \beta}$~ and H$_{\rm \gamma}$. The other stellar
parameters are not affected by this modification, as the fit to the other lines
does not change. We adopt the mass--loss rate indicated by H$_{\rm \alpha}$~ as this line
is much more sensitive to changes in this parameter and there is good
general agreement between the mass--loss rates from H$_{\rm \alpha}$~ and radio fluxes
(\cite{scuderi98}). Our result also supports this conclusion, as the
mass--loss rate derived here agrees with the upper limit of
1.5$\times$10$^{-5}$ ${\rm M}_{\odot}$ yr$^{-1}$ quoted by Bieging et al. (1989)
(1.6$\times$10$^{-5}$ if we use our values for distance and $v_{\rm \infty}$) in case the
star is a thermal emitter (the authors classify the object among the
probable thermal emitters). However, in addition to the error quoted,
\.M could be lowered by an additional 20$\%$ if we would adopt the 
value indicated by H$_{\rm \gamma}$.

The singlet \ion{He}{i} lines ($\lambda\lambda$ 4387, 4922) give a 
poor fit to the observed spectrum, partly due to the difficulties in the
continuum rectification. Therefore we have given a low weight
to these lines when determining the stellar parameters. However, since 
these lines react strongly to changes in stellar parameter, it is 
always possible to find a reasonable fit within
the error box.
The error in $\epsilon$~ is larger than for other stars and $\epsilon$\, itself is not
well constrained towards higher He abundances, because the already large He
abundance produces a saturation effect.

\begin{figure*}
{\includegraphics[width=12cm]{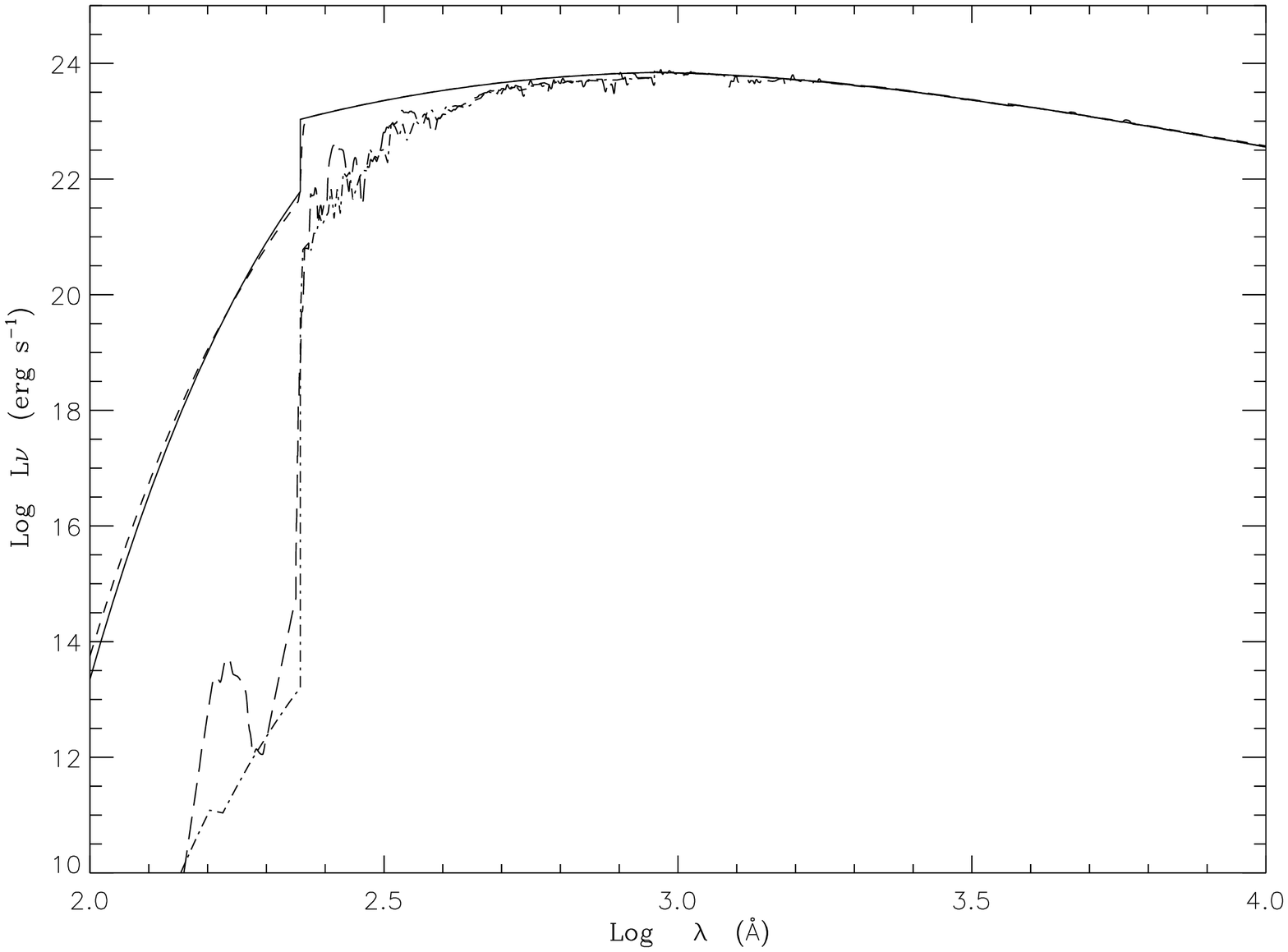}}
%{\includegraphics{figure=flujos7both.eps}}
\caption[]{Emergent energy of two models for Cyg OB2 \#7, each one
calculated both with FASTWIND and CMFGEN. The first pair of
models corresponds to the one adopted here including metal line
opacity and is represented by the lower dashed (CMFGEN) and
dash--dotted (FASTWIND) lines. The second pair corresponds to
the model adopted by Herrero et al. (2001) that did not include metal
line opacity and is represented by the upper dashed (CMFGEN)
and solid (FASTWIND) lines. We see that all
models give the same optical flux, but very different UV fluxes (see text).
We also see the good agreement between FASTWIND and CMFGEN, although
individual strong UV resonance lines are not visible in the former due to the
approximate method applied.}
\label{flux7}
\end{figure*}

The derived $T_{\rm eff}$~ is much lower than the one obtained by Herrero et al. 
(2000) using
the same code as here but without line--blocking. The derived luminosity is
also lower by more than 0.2 dex, as the radius does not change very much.
The reason can be seen in Fig.~\ref{flux7}, where we compare the emergent
energy of two models for Cyg OB2 \#7. The first model (the dash--dotted 
line in
the figure) is the one adopted here, with a $T_{\rm eff}$\, of 45\,500 K, a radius
of 14.6 $R_{\odot}$\, and metal line opacity. The second model (solid line)
corresponds to the model adopted by Herrero et al. (2001), 
with a $T_{\rm eff}$\, of 50\,000 K,
a radius of 14.8 $R_{\odot}$\, and pure H/He opacities. Both models give the same
optical luminosity and thus fit equally well the observed visual magnitude
of Cyg OB2 \#7. We additionally show in the figure the CMFGEN luminosities 
calculated with the same parameters and conditions (dashed lines).
The good agreement
supports our approximated treatment of the metal line opacity.

The reason for the similarity in derived radii is that the emergent flux is
strongly blocked in the UV by the metal line opacity and thus emerges at
higher wavelengths, including the optical. Therefore, at lower temperatures
we obtain larger optical fluxes for models that include metal line opacity
than for pure H/He models of the same temperature.

The radius needed to fit the observed visual magnitude is then significantly
smaller for models with metal line opacities (compared to unblanketed models
at the same $T_{\rm eff}$), however roughly similar to the ``old'' value derived
from unblanketed models at higher $T_{\rm eff}$. In consequence, the reduction in
luminosity is mostly due to the change in the effective temperature. Note,
however, the dramatic difference of the ionizing fluxes in the (E)UV.

The helium abundance derived here is even larger than the one obtained by
Herrero et al. (2000), 
although the error bars overlap significantly. Cyg OB2 \#7 is
thus confirmed as the only star in our sample for which we derive an
enhanced He abundance. The $\beta$ value we obtain (0.9) is slightly larger
than the one obtained from the UV analysis presented in Herrero et al. (2001),
a behaviour which has been found already in previous investigations (e.g.,
\cite{puls96}). However, our mass--loss rate is similar to the one derived 
in that work, resulting again from the fact that the optical fluxes are
similar. 

Finally, the $\log g$~ value obtained here is slightly larger than the one 
obtained by Herrero et al. (2000). 
This results in a spectroscopic mass of 39.7 ${\rm M}_{\odot}$, to be
compared with an evolutionary mass of 67.4 ${\rm M}_{\odot}$. Accounting for maximum
errors, masses of 56.7 and 59.4 ${\rm M}_{\odot}$, respectively, are possible, still not
overlapping. However, we have to remember the large He abundance derived for
Cyg OB2 \#7. This is an indication that this star might be evolved or be
affected by rotational mixing. Taken together, there is no (clear) evidence
that evolutionary and spectroscopic masses really disagree.

~\cite{wal00} and \cite{wal02} have recently studied Cyg OB2 \#7 and HD
93\,129A. They conclude, from a comparison of their spectra, that Cyg OB2
\#7 has to be cooler than HD 93\,129A, and in fact HD 93\,129A has been
reclassified as the prototype of the new O2 If$^*$ class. \cite{taresch97}
have analyzed this latter star and obtained an effective temperature of
52\,000$\pm$1\,000 K, based on the \ion{N}{v}/\ion{N}{iv}/\ion{N}{iii}
ionization equilibrium. This is similar to the Herrero et al. (2000) 
value and again
much higher than the temperature we obtain here
for Cyg OB2 \#7. Both stars display simultaneously \ion{N}{v}, \ion{N}{iv}
and \ion{N}{iii} lines in their spectra, and thus we would not expect a very
large temperature difference. Clearly, a cross--calibration of 
He and N blanketed
temperature scales for the earliest stars is an urgent task.

\paragraph{Cyg OB2 \#11}

\begin{figure*}
{\includegraphics{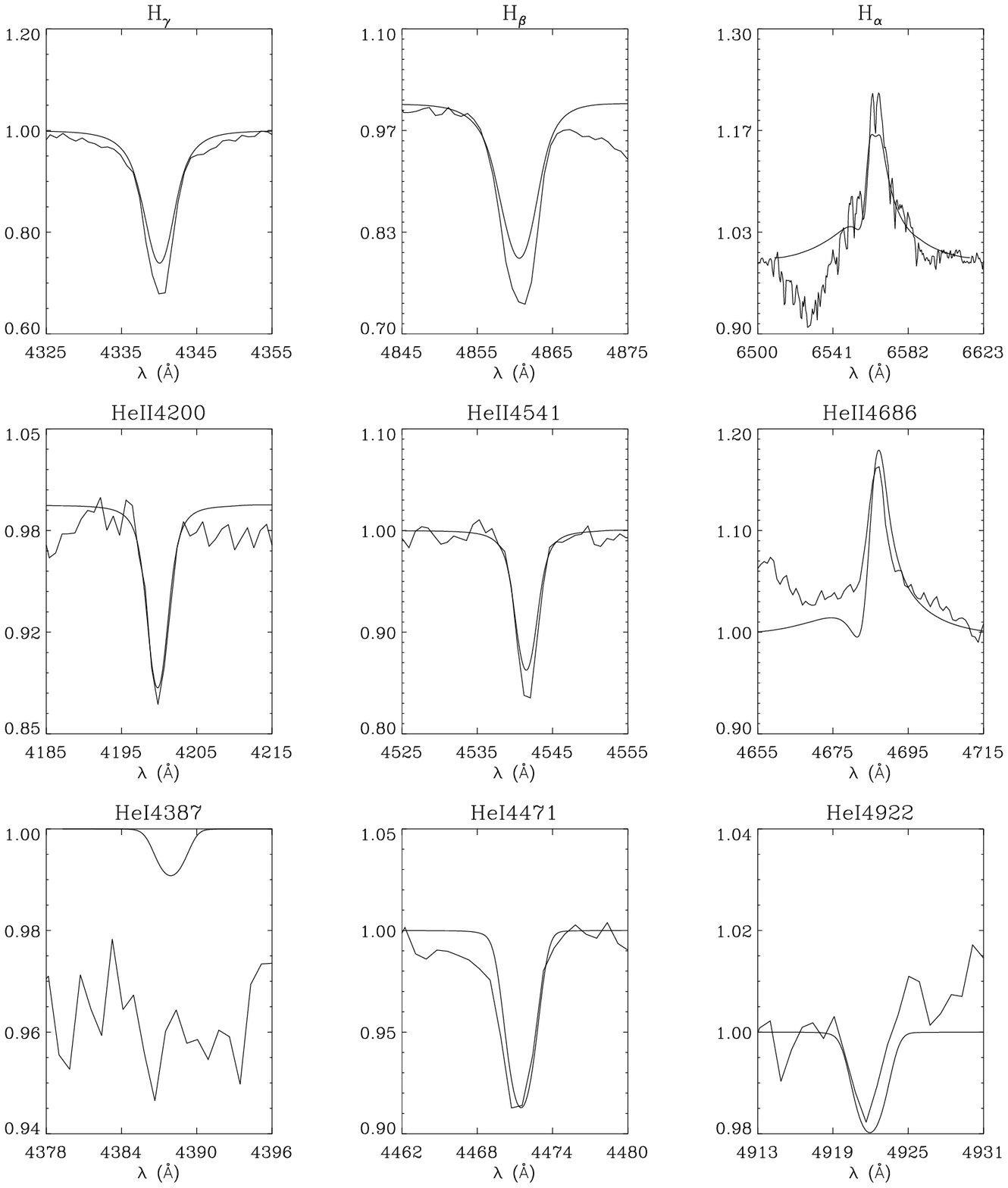}}
\caption[]{As Fig.~\ref{10lac}, however for CygOB2 \#11. See text for comments}
\label{fit11}
\end{figure*}

The fit to Cyg OB2\#11 is given in Fig.~\ref{fit11}. It shows the same
problems as the fit to Cyg OB2 \#7, 
except for the fit to the
\ion{He}{i} singlet lines, which are again affected by continuum rectification
problems. However their depths relative to the depressed local continuum
are now well predicted.

Therefore we adopt the same errors, except for
$\epsilon$, for which we adopt $\pm$0.03. It is interesting that in spite of the
extreme Of character of both stars and the similarity of the problems found,
we do not derive an enhanced He abundance for Cyg OB2 \#11. This star also
shows the same trend as Cyg OB2 \#7 towards cooler temperatures and lower
luminosities, but now the spectroscopic and evolutionary masses (73.0 and
58.1 ${\rm M}_{\odot}$, respectively) invert their usual ratio. When considering the
formal errors presented in Table~\ref{errors}, the mass ranges of both
stars overlap significantly.

The $\beta$ value we have used is again 0.9. The mass loss rate
we derive is consistent with the upper limit given by Bieging et al. (1989)
(1.4$\times$10$^{-5}$ ${\rm M}_{\odot}$ yr$^{-1}$, or 1.2$\times$10$^{-5}$ using again
our values for distance and wind terminal velocity).

\paragraph{Cyg OB2 \#8C}

\begin{figure*}
{\includegraphics{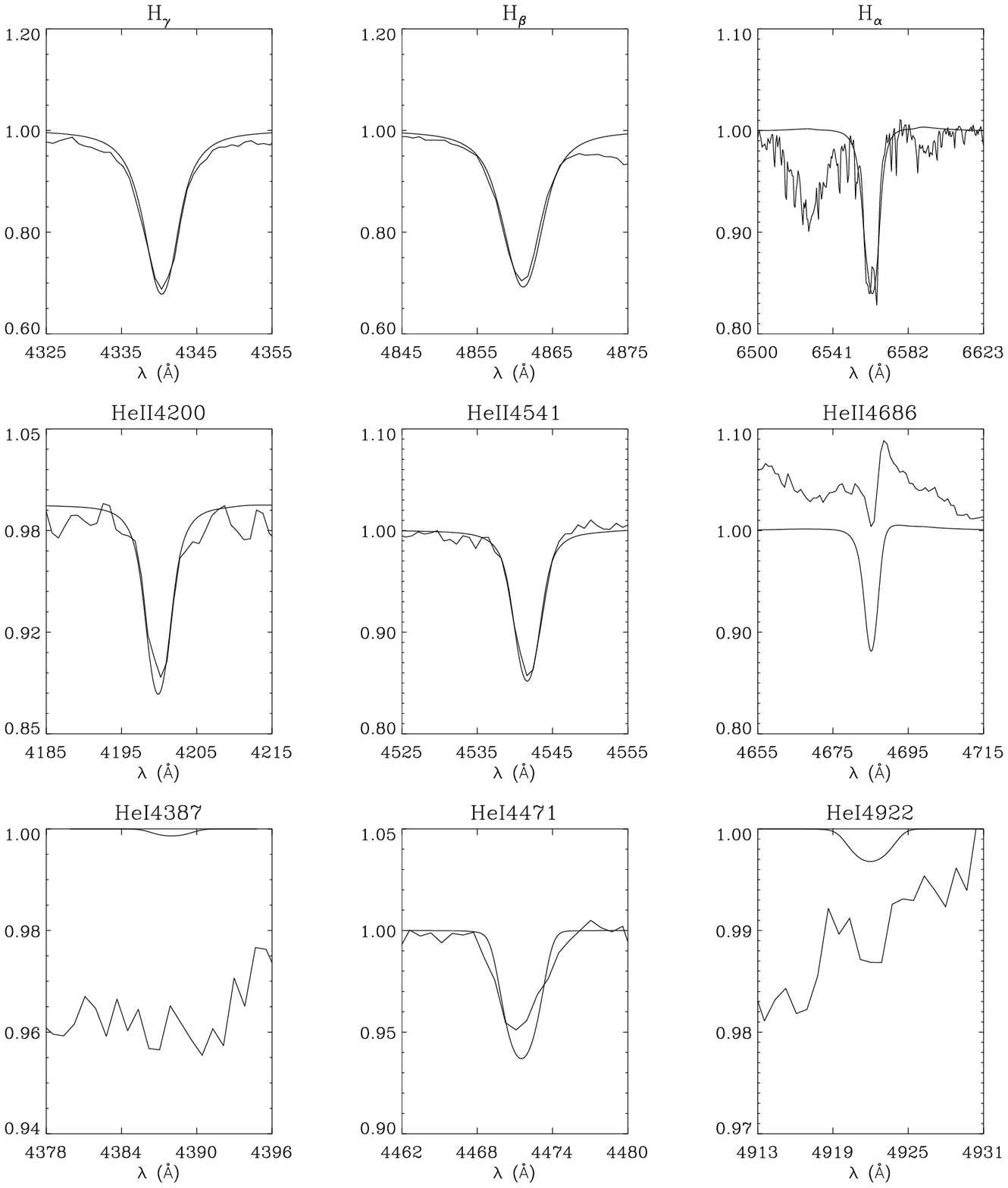}}
\caption[]{As Fig.~\ref{10lac}, however for CygOB2 \#8C. See text for comments}
\label{fit8c}
\end{figure*}

The fit to Cyg OB2\#8C is presented in Fig.~\ref{fit8c}. The only
problem is a serious failure in the predicted \ion{He}{ii} 
$\lambda$4686 line (which is not used in the fit procedure).
To fit this line one had to increase the mass--loss rate by at least
a factor of 1.7, although we note that the observed line
lies at top of a broad emission feature that we cannot reproduce.
The same comments as for Cyg OB2 \#11 apply
for the \ion{He}{i} singlet lines.

$\beta$ is not well constrained from the wings of H$_{\rm \alpha}$, as these
are in absorption. We have adopted the same value as for Cyg OB2 \#7
and \#11 (0.9), as well as the same errors.

The resulting mass--loss rate is consistent with the upper limit quoted by
Bieging et al. (1989) of 8.8$\times$10$^{-6}$ ${\rm M}_{\odot}$ yr$^{-1}$. The
gravity is large for a supergiant (a lower gravity is prohibited by the
Balmer line wings), but we find good agreement between the
spectroscopic and evolutionary masses.

\paragraph{Cyg OB2 \#8A}

\begin{figure*}
{\includegraphics{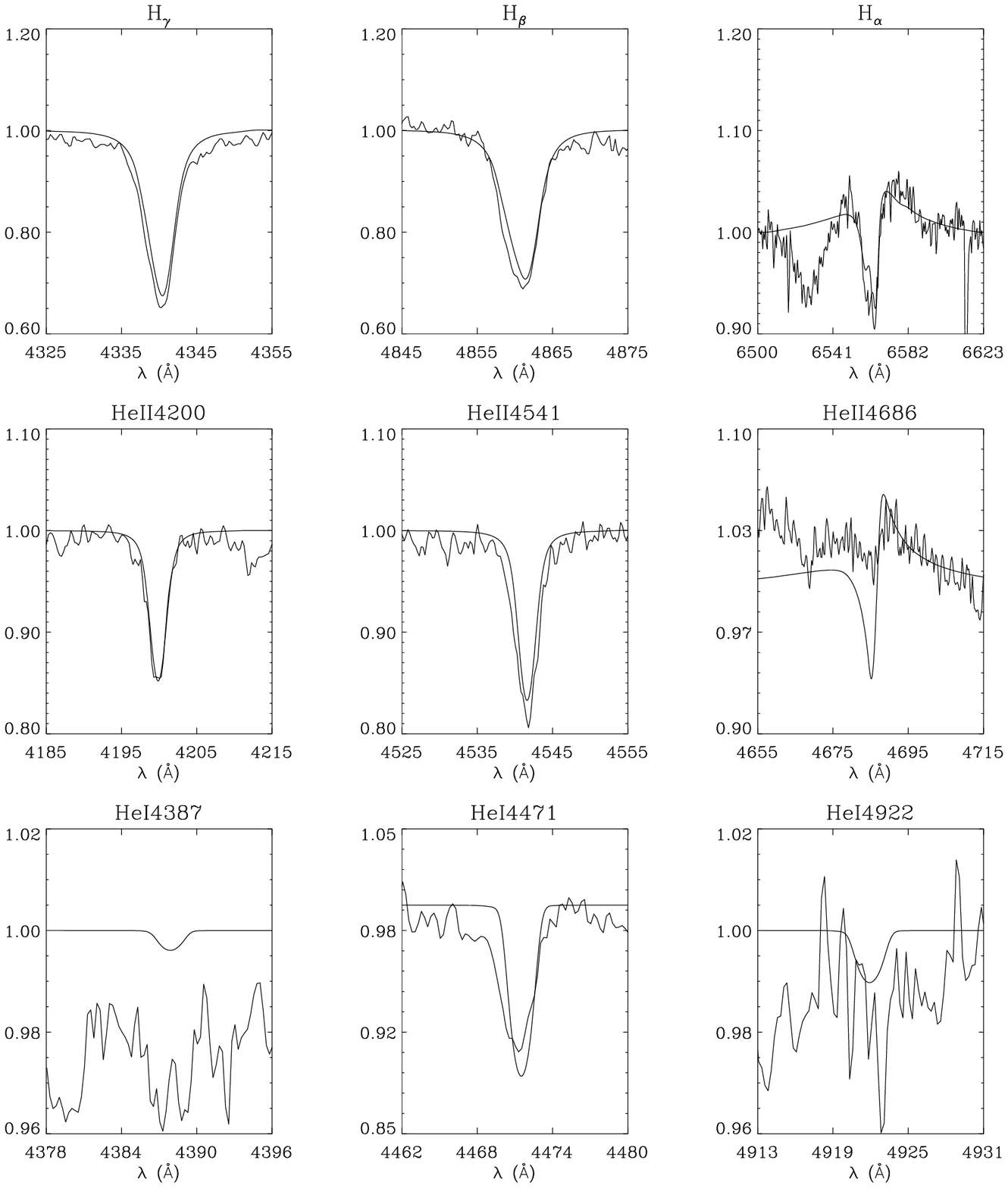}}
\caption[]{As Fig.~\ref{10lac}, however for CygOB2 \#8A. See text for comments}
\label{fit8a}
\end{figure*}

The final fit can be seen in Fig.~\ref{fit8a}. The fit to \ion{He}{ii}
$\lambda$4686 is problematic, but much less than for Cyg OB2 \#8C,
while the same comments apply for the \ion{He}{i} singlet
lines. Errors 
are the same as for Cyg OB2 \#8C. The temperature is cooler and the
luminosity lower than quoted in Herrero et al. (2001). 
However, the spectroscopic mass
is again very large (90.5 ${\rm M}_{\odot}$), larger than the evolutionary one (78.4
${\rm M}_{\odot}$), but with significant overlap when considering the errors.

The mass--loss rate we derive here is nearly a factor of two lower than the
one given in Herrero et al. (2001). 
Note, that the latter was not derived from spectrum
analysis, however was calculated from the luminosity (believed to be larger
at that time) and the Galactic WLR. 

Our new mass--loss rate in
Table~\ref{finpar} (1.35$^{+0.35}_{-0.39}\times$10$^{-5}$ ${\rm M}_{\odot}$ yr$^{-1}$)
agrees well with the radio mass--loss rate given by \cite{waldron98}
(1.97$\times$10$^{-5}$), and lies between the extreme values one would
derive from the fluxes given by Bieging et al. (1989) 
(1.1--6.1$\times$10$^{-5}$)
assuming free--free emission. Although Cyg OB2 \#8A is a known non-thermal
emitter, with variable radio flux and spectral index (Waldron et al., 1998;
\cite{bieging89}), 
our H$_{\rm \alpha}$~ mass--loss rate is of the same order of magnitude
as the radio mass--loss rates and consistent with their lower limit. This
consistency contradicts the suggestion by \cite{waldron98} that the X--ray
emission might originate from an X--ray source deeply embedded in the stellar
wind, i.e., a base corona model scenario, which would imply a much lower
mass-loss rate ($\approx$1.5$\times$10$^{-6}$ ${\rm M}_{\odot}$ yr$^{-1}$).

\paragraph{Cyg OB2 \#4}

\begin{figure*}
{\includegraphics{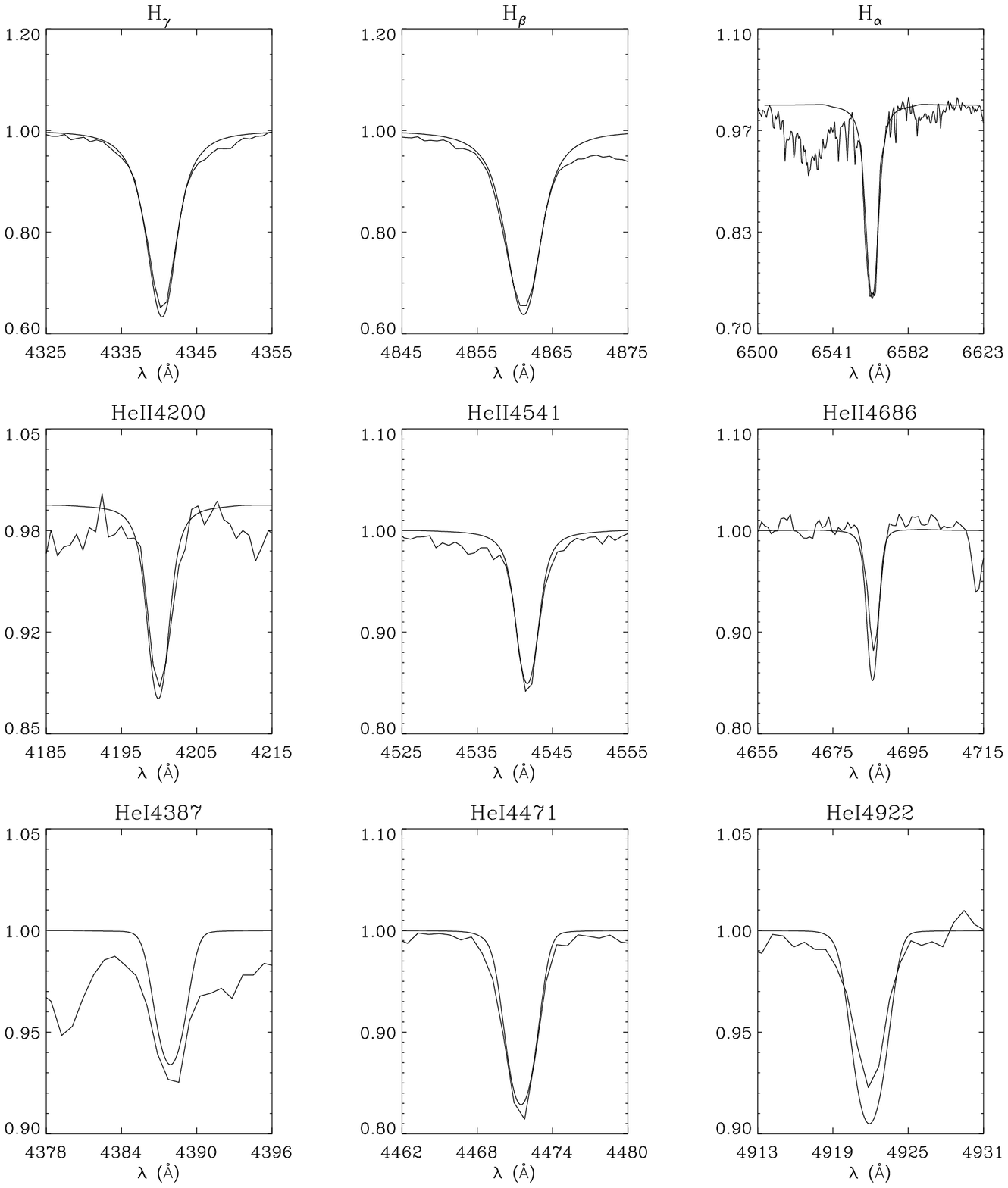}}
\caption[]{As Fig.~\ref{10lac}, however for CygOB2 \#4. See text for comments.}
\label{fit4}
\end{figure*}

The fit to Cyg OB2 \#4 is presented in Fig.~\ref{fit4}. The predicted
\ion{He}{ii} $\lambda$4686 and the singlet \ion{He}{i} line at
$\lambda$4922 are too strong in the core, 
although we note the
large scale in the corresponding plots. The fit of 
\ion{He}{i} $\lambda$4387 is acceptable taking the normalization 
into account. $\beta$ is again not well
constrained from the H$_{\rm \alpha}$\, wings and we adopt a similar value
as for the cooler stars in our sample ($\beta$= 1). However, the influence
of $\beta$ on the other stellar parameters begins to decrease and
therefore we adopt the same errors as for 10 Lac.

The mass loss rate is not well constrained towards lower values, because the
profiles react only slightly. In this case, as also for the next two Cyg OB2
stars, there are no radio mass--loss rates available to compare with
(which is an indication of a rather low value). The derived effective
temperature is still cooler than in Herrero et al. (2001), 
although the differences
begin to decrease. The evolutionary and spectroscopic mass ranges
agree within the large error bars.

\paragraph{Cyg OB2 \#10}

\begin{figure*}
{\includegraphics{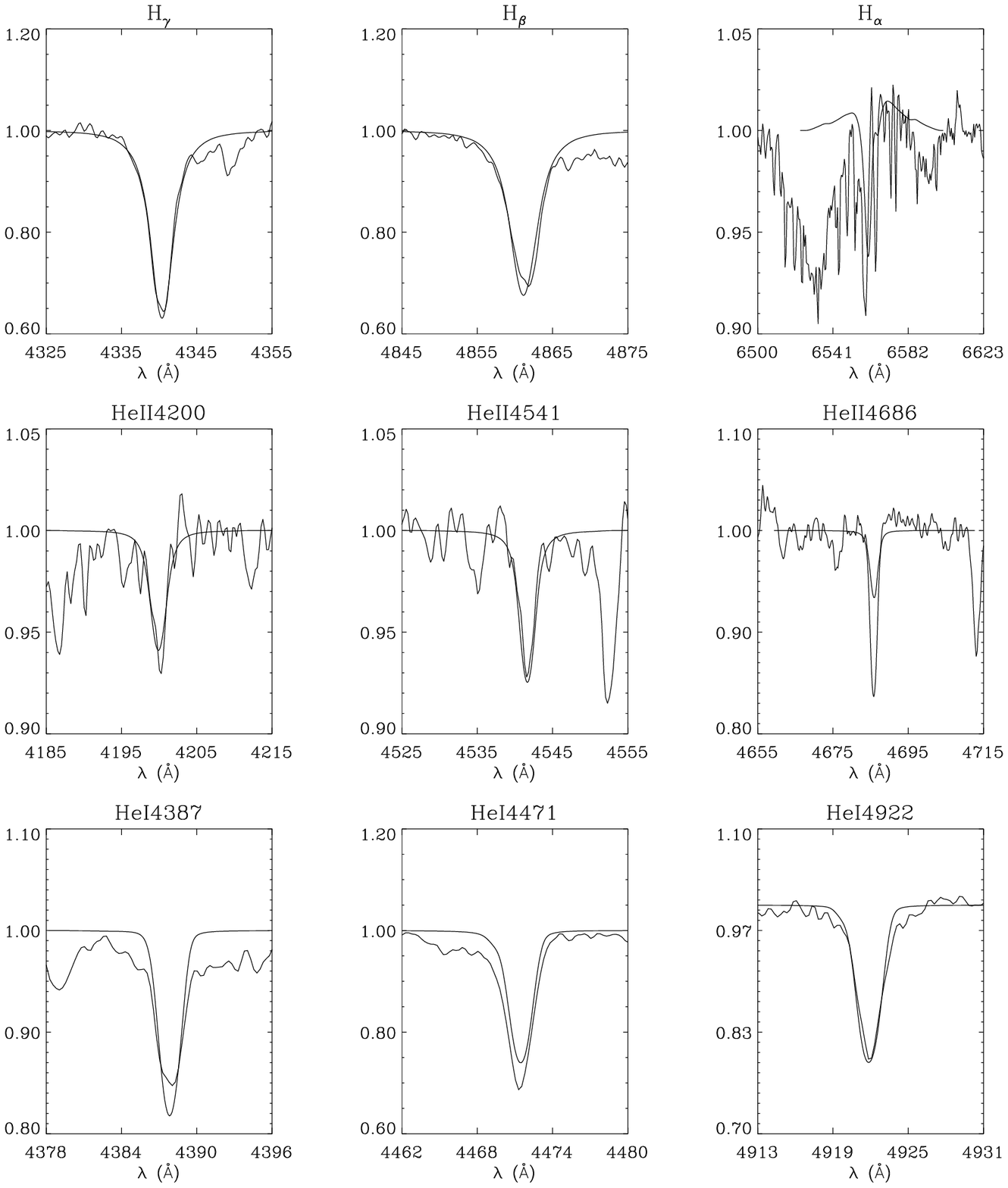}}
\caption[]{As Fig.~\ref{10lac}, however for CygOB2 \#10. See text for comments.}
\label{fit10}
\end{figure*}

The fit to Cyg OB2 \#10 is given in Fig.~\ref{fit10}. As for Cyg OB2 \#4,
the main difficulties appear in the fit of the \ion{He}{ii} $\lambda$4686
and the \ion{He}{i} lines, where from the three \ion{He}{i} lines one is
slightly too strong, the second slightly too weak and the third one fits
well. The errors adopted are the same as for Cyg OB2 \#4. The derived
effective temperature is still lower than in Herrero et al. (2001), 
but the difference
is only 2\,000 K. The spectroscopic and evolutionary masses agree well.

\paragraph{Cyg OB2 \#2}

\begin{figure*}
{\includegraphics{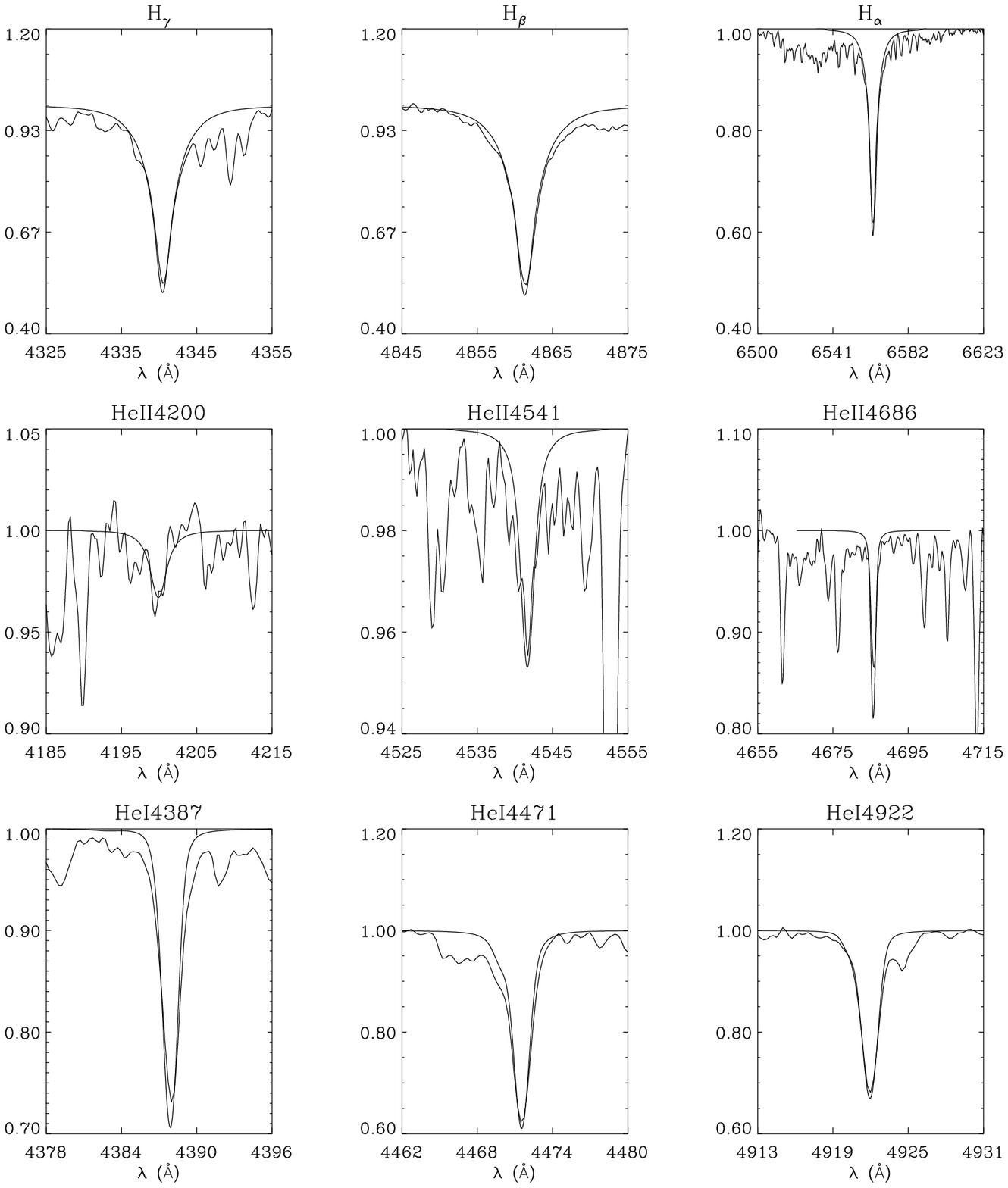}}
\caption[]{As Fig.~\ref{10lac}, however for CygOB2 \#2. See text for comments.}
\label{fit2}
\end{figure*}

The final fit to the star is shown in Fig.~\ref{fit2}. The adopted errors
are the same as for Cyg OB2 \#8C (because of the lower S/N compared to 10
Lac), although again the mass loss rate is not well constrained towards
lower values. Here, the derived effective temperature is hotter than in
Herrero et al. (2001), 
but also the derived He abundance has decreased significantly.
The masses, however, do not agree, with the spectroscopic mass being lower
than the evolutionary one by a factor of two.

Herrero et al. (2001) 
indicated some problems with the stellar classification as B1 I
and its Cyg OB2 membership, because this would result in a rather faint
absolute magnitude for its spectral class. However, the large reddening
quoted by Massey \& Thompson (1991) 
indicates that the star is probably related to or
lies beyond Cyg OB2. On the other hand, this large reddening is
comparatively low when compared to other Cyg OB2 members, which additionally
weakens the above argument. (Cyg OB2 \#2 has the fourth lowest reddening in
Table 7 of Massey \& Thompson, 1991, 
who list a total 64 Cyg OB2 stars. The stars with
the three lowest values lie in the same region of the association as Cyg OB2
\#2). Thus, we adopt the absolute magnitude derived from the canonical
distance to Cyg OB2 and assume a larger error, $\pm$0.2 instead of $\pm$0.1,
which also doubles the error in the (logarithmic) radius.

\section{Discussion}
\label{disc}

The results obtained in the preceding section allow us to address four
important aspects of the properties of massive OB supergiants: the effective
temperature scale, the ionizing fluxes, the mass discrepancy and the wind
momentum-luminosity relationship.

\subsection{The effective temperature scale for massive O supergiants}

With the inclusion of mass-loss, sphericity and line--blocking/blanketing we
derived lower temperatures than those quoted in Herrero et al. (2001), 
where we have
used plane--parallel, hydrostatic models without line--blocking for all
stars except Cyg OB2 \#7, for which a spherical model with mass--loss, but
without line--blocking was used. As we have covered various spectral types
from O3 to B1 in our analysis, we can obtain a temperature scale for O
supergiants. We note however, that Cyg OB2 \#4 is actually a luminosity
class III star, while \#2 is probably of class II.

For our earlier supergiants (O3--O7) we find temperatures that are 4\,000 to
8\,000 K cooler than in Herrero et al. (2001), 
while for the O9.5 I supergiant we
obtain 2\,000 K less and for the B1 star the temperature is now hotter, but
with a lower He abundance. These findings are in qualitative agreement with
theoretical expectations (e.g., \cite{schae94}) and with the recent
temperature scale for O dwarfs presented by Martins et al. (2002). 
These authors
obtain lower temperatures for O dwarfs by 1\,500--4\,000 K when including
line--blanketing and sphericity.

\begin{figure}
\resizebox{\hsize}{!}{\includegraphics{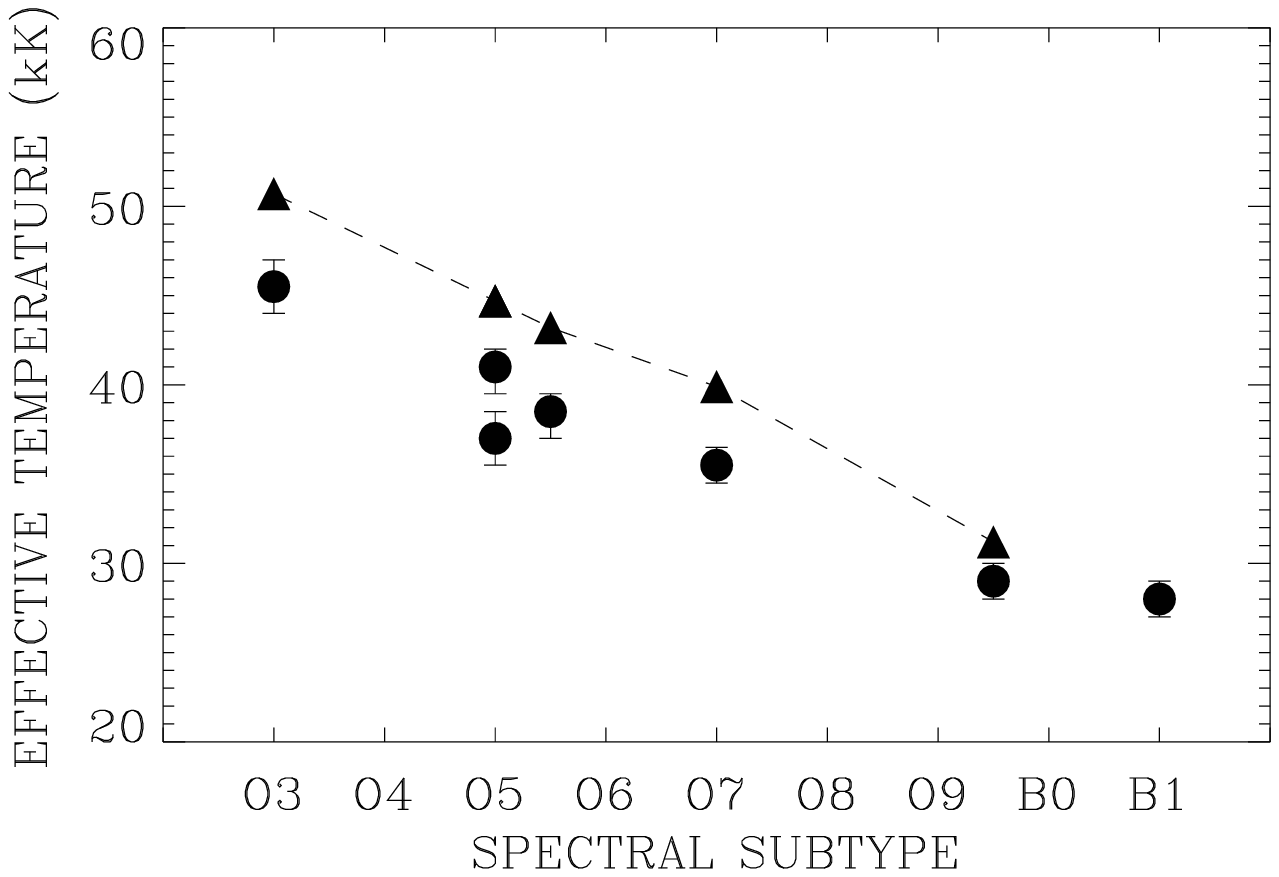}}
\caption[]{Our temperature scale for O supergiants (circles) compared
with the scale by Vacca et al. (1996) (triangles). New temperatures
are much lower, except for the relatively cool Cyg OB2 \#2 (B1I).
Note that the entry at O7 is actually a luminosity class III star.}
\label{tscale}
\end{figure}

Our temperature scale is shown in Fig.~\ref{tscale}, together with the
one from Vacca et al. (1996) for O supergiants. The effective temperature scale
from these authors has been mainly derived from analyses using pure H--He,
plane--parallel, hydrostatic models, like those performed by 
Herrero et al. (1992). It is
thus not surprising that our new scale is cooler.

In spite of the still low number statistics, we already appreciate some
interesting features in Fig.~\ref{tscale}. First, from O3 to O9 we see a
smooth temperature decline into which Cyg OB2 \#4 fits perfectly, despite its
luminosity class III. From this smoothness is excluded Cyg OB2 \#11, much
cooler than the other O5 supergiant, Cyg OB2 \#8C, and even cooler than the
O5.5 supergiant, Cyg OB2 \#8A. The main difference in their properties is
the extreme Of character of Cyg OB2 \#11, which thus appears to be related
to cooler temperatures. This is indicating to us that all spectral
signatures have a significance in terms of stellar parameters, and thus a
temperature scale using only spectral subtypes of
O supergiants, with the various nuances in their
classification scheme, will neccesarily be of limited accuracy. 

We also see that both temperature scales converge towards later spectral
types, until the B1 star, Cyg OB2 \#2. This object has a temperature that
does not seem to fit into the general behaviour, although data are still too
scarce to know whether this has any significance.

The new temperature scale and the lower luminosities will have an influence
on other aspects, e.g., on the emergent ionizing fluxes that are now much
lower than in the older models, as we have seen in the discussion of Cyg OB2
\#7.

\subsection{Ionizing fluxes from O supergiants}

Our treatment of UV metal line opacity is approximate (in the sense that we
use an approximate NLTE approach and suitably averaged lines opacities), and
we do not pretend to give a detailed description of the UV radiation field.
However, the emergent fluxes, 
should be correct in an average sense (i.e., neglecting distinct spectral
features, see again Fig.~\ref{flux7}), in particular concerning frequency
integrated quantities. (Note that the differences bluewards from 
He{\sc ii} 228 \AA~ are majorly due to different temperature
structures in the outer wind.)
Thus, we should obtain a rather correct description
of quantities extending over broad spectral regions, like the number of
photons capable of ionizing hydrogen.

This number, of course, is of extreme relevance to studies of \ion{H}{ii}
regions surrounding the stars. Vacca et al. (1992) 
have calculated ionizing fluxes
from plane--parallel, hydrostatic, LTE, line--blanketed Kurucz--models
(Kurucz, 1992) and concluded, from a comparison with more elaborate
models available at that time (\cite{schae94}), that their ionizing fluxes
should be reliable. That would indicate that line--blocking/blanketing is
the major effect when calculating the UV continuum ionizing flux {\it at a
given effective temperature}.

Our calculations seem to support this conclusion, but also indicate that for
an application to \ion{H}{ii} regions further effects have to be considered.

\begin{figure}
\resizebox{\hsize}{!}{\includegraphics{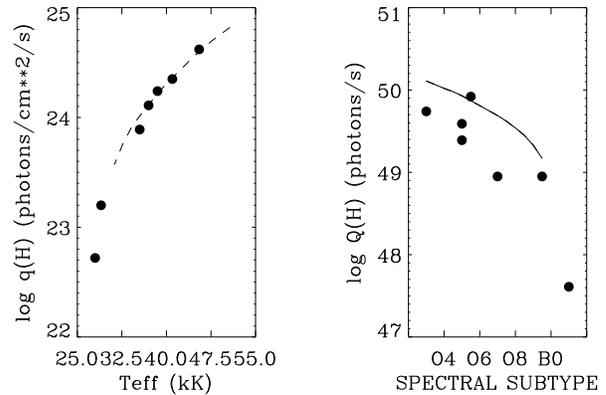}}
\caption[]{Left: ionizing fluxes (photons cm$^{-2}$ s$^{-1}$) at
the stellar surface {\it versus effective temperature}. Right: ionizing
luminosities (photons s$^{-1}$) {\it versus spectral type}. Dots correspond
to the stellar analyses presented here and lines represent the calibrations
by Vacca et al. (1996) for O supergiants.}
\label{ionf}
\end{figure}

In Fig.~\ref{ionf}a we have plotted our calculated H ionizing fluxes versus
the effective temperature of the corresponding model. The dashed line gives
the calibration by Vacca et al. (1996)
for O supergiants (their Table 7). We see
that the agreement is good. Even more, the entry departing slightly from the
relation corresponds to Cyg OB2 \#4, that has a luminosity class III. Note
that the ionizing {\em fluxes} are basically independent of stellar radius.
Thus, at given temperature the effects of metal line opacity remain the
major ingredient which detemine the emergent UV flux, and the differences  
between NLTE and LTE models seem to be small regarding the integrated
hydrogen Lyman fluxes. (This behaviour should become significantly different 
when, considering, e.g., \ion{He}{ii} ionizing fluxes because of the extreme
sensitivity of this ion to NLTE effects, in particular as function of
mass-loss.)

In Fig.~\ref{ionf}b we have plotted the H ionizing {\em luminosities} as
function {\it of spectral type}. Here, the differences with the calibration
by Vacca et al. (1996) are apparent. 
They clearly originate from our new relation
between spectral type and effective temperature (and our lower radii for
most stars, compared to the calibrations by \cite{vacca96}), i.e., the stellar
luminosities at given subtype are now smaller. (This discrepancy
between both plots, of course,
results from the former inconsistency of calibrating spectral types via
unblanketed models, however calculating the number of ionizing photons
from blanketed ones).
Furthermore, for one case (Cyg OB2 \#8A) we find that our ionizing
luminosities are {\em larger} as a consequence of the larger radius in this
particular case. Since the ionizing {\it luminosity} is the quantity which
really matters for the ionization of \ion{H}{ii} regions, our results
indicate that {\em statistically} there are fewer photons available compared
to earlier findings, but also that individual cases have to be studied in
detail, because they can depart from the general trend. 

\subsection{The mass discrepancy}

The mass discrepancy is a term that was used by Herrero et al. (1992) 
to refer to the
systematic difference between spectroscopically determined stellar masses of
OB supergiants and their evolutionary counterparts. The latter were always
systematically larger, well beyond the error bars.

\begin{figure}
\resizebox{\hsize}{!}{\includegraphics{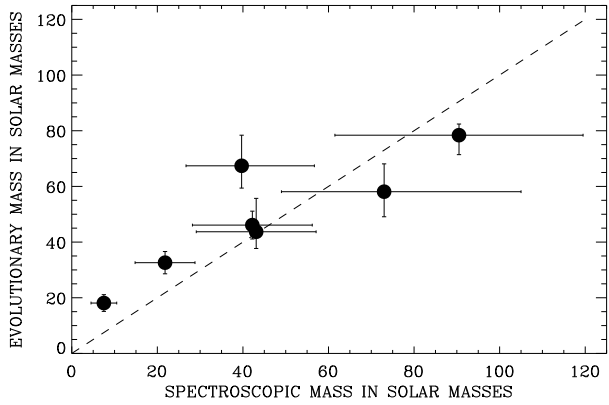}}
\caption[]{A comparison of the spectroscopic and evolutionary
masses for the Cyg OB2 supergiants. Although there are still serious
differences, no obvious systematic trend is present. See text for
details and discussion.}
\label{masses}
\end{figure}

In Fig.~\ref{masses} we have plotted the evolutionary mass (derived from the
``classical'' models by \cite{schaller92} without rotation) and the
spectroscopic masses obtained in our analysis for the Cyg OB2 supergiants.
We see that the situation is by no means satisfactory: despite the (very)
large errors adopted for the spectroscopic masses, three of the seven stars
still do not cross the one-to-one relation and for two other we find
only a marginal agreement. In fact, only two out of the seven stars have
masses that agree reasonably well. However, compared to previous diagrams
the situation seems to have improved: there is no clear systematic
trend any longer. Roughly the same number of data points lie on each 
side of the
one-to-one relation and the apparent scatter might be related to problems in
the individual analyses.

Nevertheless it is somewhat too early to conclude that the atmosphere models
with sphericity, mass--loss and metal opacity agree with the evolutionary
models without rotation, and thus give the same answer in a statistical
sense. First, the large scatter in Fig.~\ref{masses} still poses a question
for the masses of both set of models; and second, the stars in Cyg OB2
have moderate projected rotational velocities and are very young, with only
the earliest type exhibiting an enhanced He abundance. Thus, older or faster
rotating stars may have masses that disagree even more when derived using
different methods.

This result indicates that we badly need both a calibration of present mass
scales based on early type binaries (or any other reliable method), and CNO
abundances for isolated massive stars.

\subsection{The Wind momentum--Luminosity Relationship in Cyg OB2}

The original purpose of our work was to obtain a better constrained WLR for
Galactic O stars by using objects belonging to the same association, and
thus minimizing the scatter introduced by uncertainties in the relative
distances. The distance to Cyg OB2 and thus the derived luminosities may be
in error, but all determinations will be affected in a similar way.

\begin{figure}
\resizebox{\hsize}{!}{\includegraphics{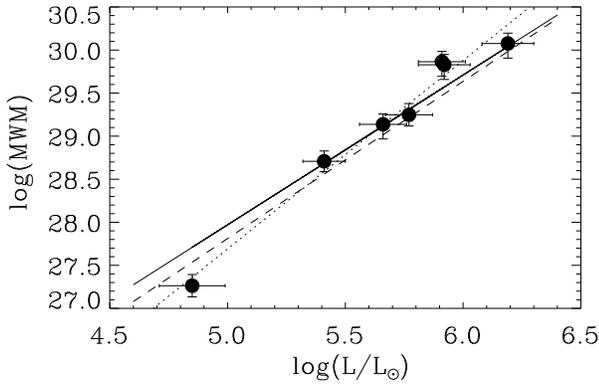}}
\caption[]{The wind momentum--luminosity relationship obtained for
the Cyg OB2 supergiants. The dotted line gives the regression obtained by
including all stars, while the solid line gives the one obtained by
including only Cyg OB2 \#8A, \#8C, \#4 and \#10. The dashed line very close
to the solid one is the theoretical relation by Vink et al. (2000). The entry
with the highest luminosity is Cyg OB2 \#8A and those close together are Cyg
OB2 \#7 and \#11, the stars with extreme Of character.}
\label{wlr}
\end{figure}

Fig.~\ref{wlr} displays the WLR obtained for our Cyg OB2 sample. Errors for
$v_{\rm \infty}$~ are taken from Herrero et al. (1991). The stars \#8A, \#8C, \#4 and \#10
follow a nice sequence with low scatter. Cyg OB2 \#7 and \#11 seem to lie
above this sequence. This is interesting because these stars display the
most extreme Of signatures in their spectra, which might be related to an
ionization change in the wind that could result in a different line-driving
force or to clumping effects that would produce an overestimation of the
mass--loss rate. Cyg OB2 \#2 seems to lie below that relation, which is
consistent with the results by Kudritzki et al. (1991), 
who found a different WLR (with
a lower offset) for the winds of early B supergiants, compared to the O-star
case. Our observed relation also agrees well with the theoretical WLR
derived by Vink et al. (2000) (the dashed line in Fig.~\ref{wlr}).

We have investigated whether a reduction in the outer minimum
electron temperature (that the model is allowed to reach, see
\cite{sph97}, Sect. 3.1) might result in lower mass--loss rates for
Cyg OB2 \#7 and \#11. While we usually assume that $T_{\rm min}$= 0.75
$T_{\rm eff}$\, (a typical value for OB stars), calculations by Pauldrach
(private communication) indicate that this value can reach values as low as
0.40 $T_{\rm eff}$\, in extreme cases. This effect has usually no influence on an
analyis as performed here, where almost all considered lines are formed in a
region with temperatures beyond this minimum. ``Only'' H$_{\rm \alpha}$\, (and
\ion{He}{ii} $\lambda$4686) might be affected in cases of extreme mass--loss,
which is the reason that we have investigated here this question.
Nevertheless, in all considered cases the resulting reduction of the derived
mass--loss rate was less than 20$\%$. Thus, even accounting for this
uncertainty we cannot bring the position of these two stars into agreement
with the WLR defined by the other four O stars.

However, taking into account the error bars, our data are still compatible
with a unique relation including all seven stars. This is also shown in
Fig.~\ref{wlr}, where we display two different regressions, one including
only Cyg OB2 \#8A, \#8C, \#4 and \#10, and one including all seven stars,
respectively. Interestingly, the relation obtained by including only the
O--{\it supergiants}, not shown for clarity, is still marginally compatible
with the position of the B--supergiant.

\begin{figure}
\resizebox{\hsize}{!}{\includegraphics{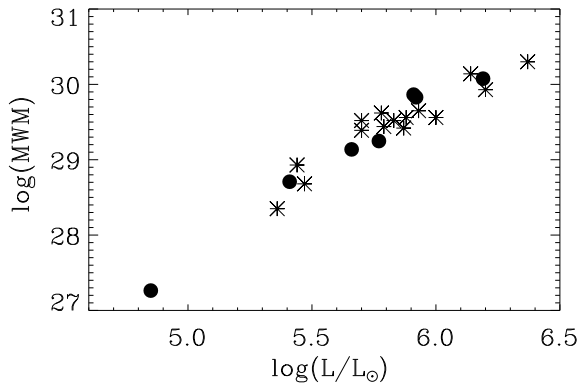}}
\caption[]{A comparison of our results for Cyg OB2 OB--supergiants
(filled dots) with the results by Puls et al. (1996) and 
Herrero et al. (2000).}
\label{wlrcomp}
\end{figure}

In Fig.~\ref{wlrcomp} we show a comparison of our data points with those
obtained by Puls et al. (1996), updated by Herrero et al. (2000) 
for some entries. We see
that both sets compare well, the most important difference being the fact
that the scatter in our data points is lower. However, at present it is not
possible to conclude whether this is a real improvement (as we have
expected) or an artefact of the low number statistics. We note that the
first possibility would imply that stars with extreme Of characteristics
follow a slightly different relation than {\it normal} O stars,
although we could not determine the reason for the relatively high position of
Cyg OB2 \#7 and \#11.

In Table~\ref{coef} we compare the coefficients we have obtained for the WLR
(by weighted least--squares fits) using different samples (see above) as well
as those quoted by Kudritzki \& Puls (2000) in their Table 2, 
and the coefficients of the
theoretical relation provided by Vink et al. (2000). 
A correct error treatment would require to take into account errors in
both axes and their correlation. This treatment is not simple and will be
presented elsewhere (\cite{markpuls02}). For the scope of the present paper
we have adopted as error an average of the errors obtained when considering
only those in the ordinate values (an underestimation) and when considering
the errors in abscissa and ordinate as uncorrelated (an overestimation).
Using both estimates for the errors, we have calculated the corresponding
regression, where the resulting values for slope and offset turned out to be
only marginally different. The final values quoted in Table~\ref{coef} have
been obtained from a straight average of these results.

We see that in spite of the visual agreement found in Fig.~\ref{wlrcomp},
our results differ significantly from those given by Kudritzki \& Puls (2000),
in particular concerning the slope of the relation.
Kudritzki \& Puls (2000) found a slope
of 1.51$\pm$0.18, while, when excluding the B supergiant, we obtain values
from 1.74$\pm$0.24 to 1.92$\pm$0.22. In addition, our values for the
vertical offset vary between 19.27 and 18.30, below the value of 20.69 of
Kudritzki \& Puls (2000).
Although the error bars allow for marginal agreement,
the conclusion is that our relation is steeper and
thus $\alpha^\prime_{\rm eff}$ (= 1/x) is smaller in our case. This indicates a
different slope of the line--strength distribution function (see Puls et al.,
2000 and \cite{kp00}, Sect. 4.1), i.e., a larger number of 
weak lines.

In contrast, we obtain very good agreement with the theoretical relation by
Vink et al. (2000).
Their relation, both in slope and offset, lies right between
our relation when considering all supergiants or only the moderate O, Of
stars, respectively. Our observations are thus compatible with a unique WLR
for all O supergiants, but favour a separation of the extreme O, Of stars.

\begin{table*}
  \caption[]{Coefficients for the wind momentum--luminosity relationship
(see eq. 2) obtained in this work and taken from Kudritzki \& Puls (2000)
and Vink et al. (2000)}

\label{coef}
    \begin{tabular}{c c c l}
      \hline 
$\log$D$_0$ & x & $\alpha^\prime_{\rm eff}$ & Comments \\
      \hline 

 16.81$\pm$1.16 & 2.18$\pm$0.21 & 0.46$\pm$0.04 & Using all Cyg OB2 stars \\
 18.30$\pm$1.82 & 1.92$\pm$0.31 & 0.52$^{+0.10}_{-0.07}$ & Cyg OB2 \#2 not included \\
 19.27$\pm$1.82 & 1.74$\pm$0.32 & 0.58$^{+0.12}_{-0.09}$  & Including only Cyg OB2 \#8A, \#8C, \#4, \#10 \\
 20.69$\pm$1.04 & 1.51$\pm$0.18 & 0.66$\pm$0.06 & From \cite{kp00} (Table2)\\
 18.68$\pm$0.26 & 1.826$\pm$0.044 & 0.548$\pm$0.013 & From \cite{vink00} (Eq. 15)
 \end{tabular}
\end{table*}

\section{Conclusions}
\label{conc}

We have analyzed a homogeneous set of spectra of Cyg OB2 supergiants with
spectral types ranging from O3 to B1. The analysis has been performed using
a new version of the code presented by Santolaya--Rey et al. (1997) 
that includes an
approximate treatement of metal line--blocking/blanketing. A test applied
to the O9 V star 10 Lac resulted in an excellent fit, at temperatures lower
than those obtained previously (\cite{h92}) and in agreement with other
findings (\cite{hub98}; \cite{martins02}).

The fits to the supergiants are also satisfactory, except for the
discrepancy between the fits of H$_{\rm \alpha}$\, and other higher Balmer lines for
stars with dense winds (see discussion in Sect. 6.1 of \cite{h00})
and the pure absorption profile predicted for \ion{He}{ii} $\lambda$
4686 in Cyg OB2 \#8C instead of the P-Cygni profile observed.

We obtain temperatures that are cooler than in our previous analyses,
as a consequence of the effects of sphericity, mass--loss and metal--line
blocking/blanketing. Thanks to our coverage in spectral type, we can
construct a temperature scale for O--supergiants. This temperature
scale is 4\,000 to 8\,000 K cooler at early types than the one presented
by Vacca et al. (1996), mainly based on pure H-He analyses. An important
additional result is that at the same spectral subtype, Cyg OB2 \#8C
and \#11 have a difference of 3\,000 K in their effective temperatures,
reflecting their very different mass--loss rates. For the hotter stars,
additional information based on a nitrogen temperature scale is needed,
as the He lines become progressively either insensitive to changes in stellar
parameters or simply too weak. 

Models including metal line opacities result in cooler temperatures and
similar radii than those using pure H/He opacities, therefore giving smaller
luminosities. Together with the blocking produced in the (E)UV, the
corresponding ionizing luminosities may change drastically, which will have
a large impact on studies of the surrounding regions. However, each
particular case has to be treated individually.

The results obtained indicate that Cyg OB2 stars are very massive.
Only for two of the seven stars we found spectroscopic 
or evolutionary masses below 40 ${\rm M}_{\odot}$. Although there are still significant
differences between spectroscopic and evolutionary masses, 
we do not find any obvious systematic pattern. Thus, at present we cannot
conclude that there is a discrepancy between both sets of masses in Cyg OB2.
However, this association is a very young one and the stars
analyzed show only moderate projected rotational velocities: 
it might be possible that rotationally induced effects simply have had 
no sufficient time to become apparent. It is remarkable that we
found an enhanced He abundance for only one of the stars, Cyg OB2 \#7.

Finally, we derived a new calibration of the wind momentum--luminosity
relationship for O supergiants, including errors resulting from our
analysis. Our data indicate that there might be a different relation for
extreme Of and for {\it moderate} O, Of stars, respectively, perhaps
indicating different ionization conditions or clumping in the wind.
Considering only the {\it moderate} O, Of stars, we obtain a very low scatter
in the relation, but this might reflect only the low number statistics. In fact,
our still limited sample is also consistent with a unique relation including
all stars, even the B supergiant. Clearly, more data are needed to
disentangle whether the large modified wind momentum rate we derive for the
extreme Of stars is a real, physical effect or just the result of our poor
statistics.

\acknowledgements{We thank an annonymous referee for very valuable 
suggestions that helped to improve the paper. FNP thanks the spanish
MCyT for grants ESP98-1351 and PNAYA2000-1784 within the Ram\'on
y Cajal Program. AHD thanks support
from the spanish MCyT under project PNAYA2001--0436.}


\begin{thebibliography}{}

\bibitem[Abbott \& Lucy, 1985]{AbbLu85}
Abbott, D.C. \& Lucy, L.B. 1985, ApJ 288, 679
\bibitem[Bianchi \& Garc\'{\i}a, 2002]{bianchi02}
Bianchi, L. \& Garc\'{\i}a, M. 2002, ApJ, in press
\bibitem[Bieging et al., 1989]{bieging89}
Bieging, J.H., Abbott, D.C., \& Churchwell, E.B. 1989, ApJ 340, 518
\bibitem[Bresolin et al., 2002]{bre02}
Bresolin, F., Kudritzki, R.P., Lennon D.J. et al. 2002, ApJ, in press
\bibitem[Bohlin, 1975]{boh75}
Bohlin, R.C. 1975, ApJ 200, 402
\bibitem[Conti \& Alschuler, 1971]{conti71}
Conti, P.S. \& Alschuler, W.R. 1971, ApJ 170, 325
\bibitem[Crowther \& Bohannan, 1997]{crow97}
Crowther, P.A. \& Bohannan, B. 1997, A\&A 317, 532
\bibitem[Crowther et al., 2002]{crow02}
Crowther, P.A., Hillier, D.J., \& Evans C.J. et al. 2002, ApJ, in press
\bibitem[Friend \& Abbott, 1986]{FA86}
Friend, D.B. \& Abbott, D.C. 1986, ApJ 311, 701
\bibitem[Fullerton et al., 2000]{full00}
Fullerton, A.W., Crowther, P.A., \& De Marco O. et al. 2000, ApJ 538, L43
\bibitem[Groenewegen \& Lamers, 1989]{gl89}
Groenewegen, M.A.T. \& Lamers H.J.G.L.M. 1989, A\&AS 79, 359
\bibitem[Hamann, 1981]{ham81}
Hamann, W.R. 1981, A\&A 93, 353
\bibitem[Haser, 1995]{has95}
Haser, S.M. 1995, PhD Thesis, Universit\"ats--Sternwarte der
Ludwig--Maximillian Universit\"at, M\"unchen
\bibitem[Herrero et al., 1992]{h92}
Herrero, A., Kudritzki, R.P., \& V\'{\i}lchez, J.M. et al. 1992, A\&A 261, 209
\bibitem[Herrero et al., 1999]{h99}
Herrero, A., Corral, L.J., Villamariz, M.R., \& Mart\'{\i}n, E.L.
1999, A\&A 348, 542
\bibitem[Herrero et al., 2001]{h01}
Herrero, A., Puls, J., \& Corral, L.J. et al. 2001, A\&A 366, 623
\bibitem[Herrero et al., 2000]{h00}
Herrero, A., Puls, J., \& Villamariz, M.R. 2000, A\&A 354, 193
\bibitem[Hillier \& Miller, 1998]{hill98}
Hillier, D.J. \& Miller, D.L. 1998, ApJ 496, 407
\bibitem[Howarth \& Prinja, 1989]{howp89}
Howarth, I.D. \& Prinja, R.K. 1989, ApjS 69, 527
\bibitem[Hubeny et al., 1998]{hub98}
Hubeny, I., Heap, S.R., \& Lanz, T. 1998, ASP Conf. Series Vol. 131, 108
\bibitem[Hubeny \& Lanz, 1995]{hub95}
Hubeny, I. \& Lanz, T. 1995, ApJ 439, 875
\bibitem[Jenkins, 1970]{jen70}
Jenkins, E.B. 1970, in IAU Symp. 36 on Ultraviolet Stellar Spectra
and Ground Based Observations, L. Houziaux \& H.E. Butler eds.,
(Dordrecht: Reidel), p. 281
\bibitem[Kn\"odlseder, 2000]{knod00}
Kn\"odlseder, J. 2000, A\&A 360, 539 
\bibitem[Kudritzki \& Puls, 2000]{kp00}
Kudritzki, R.P. \& Puls, J. 2000, ARA\&A 38, 613
\bibitem[Kudritzki et al., 1999]{kud99}
Kudritzki, R.P., Puls, J., \& Lennon, D.J. et al. 1999, A\&A 350, 970
\bibitem[Kudritzki et al., 1989]{kud89}
Kudritzki, R.P., Pauldrach, A.W.A., Puls, J., \& Abbott, D.C.
1989, A\&A 219, 205
\bibitem[Kurucz (1992)]{kurucz92}
Kurucz, R.L. 1992, in IAU Symp. 149, The Stellar Populations of Galaxies,
ed. B. Barbuy \& A. Renzini, Kluwer, Dordrecht, p. 225 
\bibitem[Lamers et al., 1999]{lam99}
Lamers, H.J.G.L.M., Haser, S., de Koter, A., \& Leitherer, C. 
1999, ApJ 516, 872
\bibitem[Leitherer, 1982]{leith82}
Leitherer, C., Hefele, H., Stahl, O., \& Wolf, B., 1982, A\&A 108, 102
\bibitem[Markova \& Puls, 2002]{markpuls02}
Markova, N. \& Puls, J. 2002, in prep.
\bibitem[Martins et al., 2002]{martins02}
Martins, F., Schaerer, D., \& Hillier, D.J. 2002, A\&A 382, 999
\bibitem[Massey \& Thompson, 1991]{masth91}
Massey, P. \& Thompson, A.B. 1991, AJ 101, 1408
\bibitem[Mathys, 1988]{mathys88}
Mathys, G. 1988, A\&AS 76, 427
\bibitem[McCarthy et al., 1997]{mc97}
McCarthy, J.K., Kudritzki, R.P., \& Lennon, D.J. et al. 1997, ApJ 482, 757 
\bibitem[Meynet \& Maeder, 2000]{mey00}
Meynet, G. \& Maeder, A. 2000, A\&A 361, 101
\bibitem[Najarro, 2002]{naj02}
Najarro, F. 2002, in prep.
\bibitem[1998]{Pauldrachetal98}
Pauldrach, A.W.A., Lennon, M., \& Hoffmann T.L. et al. 1998, PASPC 131, 258
\bibitem[Pauldrach et al., 2001]{paul01}
Pauldrach, A.W.A., Hoffmann, T.L., \& Lennon M. 2001, A\&A 375, 161
\bibitem[Puls, 1991]{puls91}
Puls, J. 1991, A\&A 248, 581
\bibitem[Puls, 2002]{puls02}
Puls, J. 2002, in prep.
\bibitem[Puls et al., 1993]{pof93}
Puls, J., Owocki, S.P., \& Fullerton, A.W. 1993, A\&A 279, 457
\bibitem[Puls et al., 1996]{puls96}
Puls, J., Kudritzki, R.P., \& Herrero, A. et al. 1996, A\&A 305, 171
\bibitem[2000]{Pulsetal00}
Puls, J., Springmann, U., \& Lennon M. 2000, A\&AS 141, 23
\bibitem[Santolaya--Rey et al., 1997]{sph97}
Santolaya--Rey, A.E., Puls, J., \& Herrero A. 1997, A\&A 323, 488
\bibitem[Schaerer \& Schmutz, 1994]{schae94}
Schaerer, D. \& Schmutz, W. 1994, A\&A 288, 231
\bibitem[Schaller et al. (1992)]{schaller92}
Schaller, G., Schaerer, D., Meynet, G., \& Maeder, A. 1992, A\&AS 96, 269
\bibitem[1991]{Schmutz91}
Schmutz, W. 1991, in: Stellar Atmospheres: Beyond Classical Models, eds.
 L. Crivellari, I. Hubeny and D.G. Hummer, NATO ASI Series C Vol. 341,
 Kluwer, Dordrecht, p. 191
\bibitem[Schulte, 1958]{schul58}
Schulte, D.H. 1958, ApJ 128, 41
\bibitem[Shull \& Van Steenberg, 1985]{shu85}
Shull, J.M. \& Van Steenberg, M.E. 1985, ApJ 294, 599
\bibitem[Scuderi et al., 1998]{scuderi98}
Scuderi, S., Panagia, N., Stanghellini, C., Trigilio, C., \& Umana, G. 
1998, A\&A 332, 251
\bibitem[Taresch et al. (1997)]{taresch97}
Taresch, G., Kudritzki, R.P., \& Hurwitz, M. et al. 1997, A\&A 321, 531
\bibitem[Urbaneja et al., 2002]{urb02}
Urbaneja, M.A., Herrero, A., \& Kudritzki, R.P. et al. 2002, A\&A 386, 1019
\bibitem[Vacca et al., 1996]{vacca96}
Vacca, W.D., Garmany, C.D., \& Shull J.M. 1996, ApJ 460, 914
\bibitem[Villamariz et al., 2002]{villa02}
Villamariz, M.R., Herrero, A., Butler, K., \& Becker, S.R. 2002, A\&A 388, 940
\bibitem[Vink et al., 2000]{vink00}
Vink, J.S., de Koter, A., \& Lamers, H.J.G.L.M. 2000, A\&A 362, 295
\bibitem[Walborn, 1973]{wal73}
Walborn, N.R. 1973, AJ 78, 1067
\bibitem[Walborn et al. (2002)]{wal02}
Walborn, N.R., Howarth, I.D., \& Lennon, D.J. et al. 2002, AJ 123, 2754
\bibitem[Walborn et al. (2000)]{wal00}
Walborn, N.R., Lennon, D.J., \& Heap, S.R. et al. 2000, PASP 112, 1243
\bibitem[Walborn, 1985]{wal85}
Walborn, N.R., Nichols--Bohlin, J., \& Panek, R.J. 1985, NASA Ref. Pub. 1155
\bibitem[Waldron et al. (1998)]{waldron98}
Waldron, W.L., Corcoran, M.F., Drake, S.A., \& Smale, A.P. 1998, ApJS 118, 217
 \end{thebibliography}
\end{document}